\documentclass[journal]{IEEEtran}
\usepackage{amsmath,amsfonts}
\usepackage{algorithmic}
\usepackage{algorithm}
\usepackage{array}
\usepackage[caption=false,font=normalsize,labelfont=sf,textfont=sf]{subfig}
\usepackage{textcomp}
\usepackage{stfloats}
\usepackage{url}
\usepackage{verbatim}
\usepackage{graphicx}
\usepackage{cite}
\usepackage{booktabs}
\usepackage{multirow}
\usepackage{graphicx}
\usepackage{refcount}
\begin{document}


\title{FineSpeaker: Reliable Speaker Representation Learning at Fine Temporal Resolution}
\author{
Yongkang Yin$^{1,2, \dagger}$, Wenhao Yang$^{3}$, Lichun Fan$^{3}$, Jian Luan$^{3}$, Hang Su$^{3*}$, Yuexian Zou$^{1,2*}$\\[2pt]
$^{1}$\textit{Guangdong Provincial Key Laboratory of Ultra High Definition lmmersive Media Technology, Shenzhen Graduate School, Peking University}\\
$^{2}$\textit{ADSPLAB, School of ECE, Peking University, Shenzhen, China}\\
$^{3}$\textit{MiML Plus, Xiaomi Inc., Beijing, China}\\
\texttt{\{yinyongkang@stu.pku.edu.cn, suhang10@xiaomi.com, zouyx@pku.edu.cn\}}\\
\thanks{$^\dagger$ This work was conducted during an internship at Xiaomi Inc.}
\thanks{$^*$ Corresponding authors.}
}



\maketitle
\begin{abstract}
Speaker representation learning is fundamental to speech processing tasks such as speaker verification and diarization.
Although long-context speaker encoders can produce stable and discriminative embeddings, short-duration verification and diarization require finer-grained temporal representations. Short temporal contexts limit the available speaker evidence and make speaker identity difficult to disentangle from phonetic and acoustic variability. This creates a temporal resolution–representation reliability trade-off: speaker representations should maintain local within-speaker consistency, between-speaker discriminability and remain sensitive to speaker changes at fine temporal resolution.
Accordingly, we propose a framework for reliable speaker representation learning at fine temporal resolution. FineSpeaker first constructs resolution-controllable frame-wise speaker representations through mean pooling over an adjustable $n$-frame window, enabling flexible control of representation granularity.
It then introduces a speaker encoder with local temporal modeling that exploits short-range speaker context to improve the stability and discriminability of short-duration and frame-wise representations. Furthermore, we develop a generative disentanglement strategy for optimizing frame-level speaker representations, combining VAE-based probabilistic modeling with flow matching to characterize speaker-related variation and suppress non-speaker interference.
Experiments demonstrate competitive full-utterance performance, improved fine-grained speaker verification on VoxCeleb and CN-Celeb, and strong zero-shot diarization results, highlighting the benefits of balancing temporal resolution and representation reliability.
\end{abstract}


\section{Introduction}
\IEEEPARstart{S}{peaker} representation learning aims to extract discriminative features that capture speaker identity from speech, and is widely used in speaker verification~\cite{wang2025review} and speaker diarization~\cite{park2022review}. However, most existing methods are designed to learn representations from relatively long speech segments, whereas many practical scenarios require reliable speaker modeling from short speech segments. In speaker verification, the available speech may be limited in duration. In speaker diarization, representations with higher temporal resolution are needed to track speaker identities and changes over time. Fine-grained speaker representations are therefore important for reliable verification under limited speech and can simplify multi-stage diarization pipelines while supporting more precise temporal speaker modeling.
\IEEEpubidadjcol
\begin{figure}[t]
    \centering
    \includegraphics[width=\columnwidth]{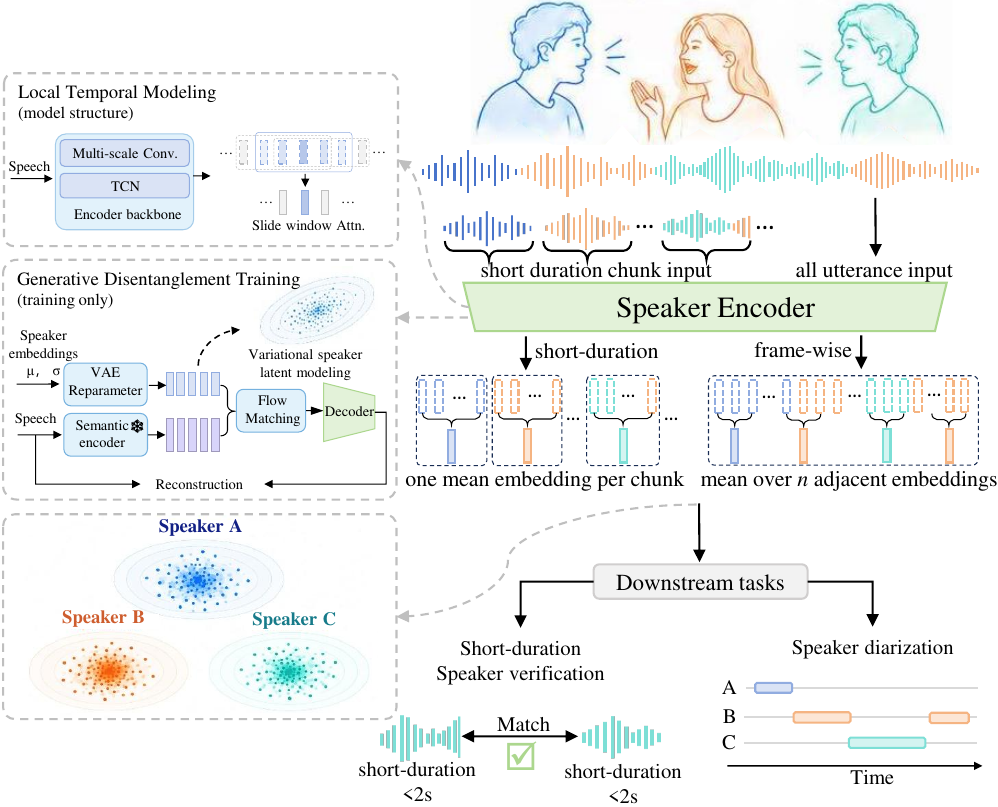}
    \caption{Overview of the proposed fine-grained speaker representation learning method, including the speaker encoder architecture, the generative disentanglement training scheme, and multi-granularity representation extraction. And evaluated on downstream speaker verification and diarization tasks.}
    \label{fig:work}
\end{figure}

Most existing speaker representation learning methods are designed for long-duration utterances and aim to learn stable and discriminative utterance-level speaker embeddings~\cite{snyder2018x,desplanques2020ecapa,yu2021cam,wang2023cam++,li2017deep, zhou2021resnext, chen2023enhanced,chen2024eres2netv2, yakovlev2024reshape,yakovlev2026redimnet2}. To improve these embeddings, supervised systems commonly use multi-scale feature extraction, cross-layer feature fusion, and attention mechanisms. Self-supervised methods learn speaker-discriminative features through contrastive learning~\cite{zhang2021contrastive}, Moco-based~\cite{xia2021self}, DINO-based~\cite{cho2022non}, and self-distillation~\cite{li2026enhancing}. Large pre-trained speech models have also been adapted to speaker recognition through task-specific fine-tuning and layer-wise feature fusion~\cite{vaessen2022fine, radford2023robust, zhao2024whisper, yin2026whisperdiari, han2025leveraging}. In addition, generative and disentanglement-based methods separate speaker information from speech content to reduce the influence of non-speaker factors. Also short-duration speaker modeling has been explored using multi-scale feature fusion, local temporal modeling, and combinations of pre-trained speech features with multi-resolution encoders~\cite{jung2020improving, mun2023frequency,kim2022rawnext,chen2024eres2netv2,deng2026src,myoung2026short}. These studies extend speaker representation learning to shorter speech segments and provide improved utterance-level embeddings under limited-duration conditions.

Despite recent advances, learning reliable fine-grained speaker representations remains challenging.
First, existing pooling methods do not effectively balance estimation reliability and temporal flexibility. Within single-speaker regions, shorter windows typically provide fewer effective samples, making identity estimates less stable. With few frames, second-order statistics may be less reliable, while attention weights may be sensitive to phonetic and acoustic changes. Moreover, utterance-level pooling usually produces a single embedding, without flexible control over local window sizes and output rates.
Second, temporal modeling does not always use context at a suitable scale. Individual frames contain limited speaker information, and pooling alone does not ensure effective modeling of temporal relationships between neighboring frames. Longer context can provide more speaker evidence, but combining information across speaker boundaries may introduce cross-speaker mixing. Local temporal modeling is therefore needed to improve local consistency and speaker discriminability while reducing cross-speaker interference.
Third, training objectives applied mainly to pooled embeddings provide only indirect constraints on frame-level features. Models can distinguish speakers by combining information across frames, even when individual features remain sensitive to speech content and acoustic conditions. These non-speaker factors may increase within-speaker variation when speech is limited. This motivates frame-level distribution modeling and constraints to characterize speaker-related variation and help separate it from short-term nuisance factors. 
These limit short-duration speaker verification and make speaker diarization more dependent on complex preprocessing~\cite{bredin2023pyannote,chen20253d}.

We aim to learn stable and discriminative fine-grained speaker representations. To this end, we propose FineSpeaker, a framework that addresses the temporal resolution–stability trade-off in fine-grained speaker representation learning by enforcing local speaker consistency and disentangling speaker identity from short-term non-speaker variations. The main contributions are as follows:
\begin{itemize}
\item \textbf{Frame-wise speaker representation scheme.} We introduce a resolution-controllable scheme that produces frame-wise speaker representations through local mean pooling over $n$ neighboring frame-level features, with $n$ controlling the representation rate.
\item \textbf{Local temporal speaker modeling.} We strengthen local speaker evidence using TCN-augmented convolutional backbone followed by sliding-window attention, promoting locally consistent and discriminative fine-grained representations while limiting cross-speaker interference.
\item \textbf{Generative disentanglement training.} We develop a frame-level generative disentanglement mechanism that combines VAE-based probabilistic latent modeling with flow-matching constraints to separate speaker-related variation from short-term nuisance factors.
\item \textbf{Evaluation across tasks and temporal granularities.}
FineSpeaker achieves competitive full-utterance verification performance and outperforms the evaluated baselines under short-duration and frame-wise conditions on VoxCeleb and CN-Celeb. Zero-shot diarization evaluations on English and Chinese datasets further demonstrate its effectiveness across temporal granularities.
\end{itemize}

\section{Background and Related Work}
\subsection{Speaker Representation Learning}
Speaker representation learning is a fundamental component of modern
speaker verification systems, aiming to extract speaker-discriminative
features from speech. A typical model structure consists of a frame-level feature extraction backbone, a temporal pooling module, and an embedding projection layer. Given an speech feature sequence $\mathbf{X}\in\mathbb{R}^{T\times F}$, the backbone $B(\cdot)$ extracts a frame-level feature sequence $\mathbf{H}$. The pooling module $P(\cdot)$ aggregates the frame-level features, and
the projection layer $F(\cdot)$ maps the pooled feature to a fixed-dimensional utterance-level speaker embedding as Eq.~\eqref{eq:speaker_embedding}:
\begin{equation}
\mathbf{H}=B(\mathbf{X}), \qquad
\mathbf{x}_{\mathrm{utt}}=F\big(P(\mathbf{H})\big).
\label{eq:speaker_embedding}
\end{equation}
where
$\mathbf{H}\in\mathbb{R}^{T\times D}$ denotes the frame-level feature
sequence and
$\mathbf{x}_{\mathrm{utt}}\in\mathbb{R}^{D_{\mathrm{emb}}}$ denotes utterance-level speaker embedding.

Speaker encoder structures commonly use TDNN- or ResNet-family backbones, including x-vector~\cite{snyder2018x}, ECAPA-TDNN~\cite{desplanques2020ecapa}, CAM++~\cite{yu2021cam,wang2023cam++}, and ResNet/Res2Net variants ~\cite{li2017deep,zhou2021resnext, chen2023enhanced, chen2024eres2netv2}. These models improve speaker representation learning through residual connections, multi-scale feature extraction, and attention mechanisms, while recent methods such as ReDimNet~\cite{yakovlev2024reshape,yakovlev2026redimnet2} further model temporal dependencies. And temporal pooling module usually relies on statistics pooling, attentive statistics pooling, or their variants, which aggregate first- and second-order statistics, such as the mean and standard deviation, over multiple frames to produce utterance-level speaker embeddings.

Short-duration speech remains challenging for speaker representation learning. Existing methods mainly address this problem through multi-scale feature aggregation, specialized architectures, and pre-trained speech models. Feature Pyramid Modules combine features from different network levels to capture multi-scale speaker information and improve robustness to variable-duration inputs ~\cite{jung2020improving,mun2023frequency}. RawNeXt~\cite{kim2022rawnext} uses deep layer aggregation and dynamic scaling to handle inputs with different durations, while Frequency and Multi-Scale Selective Kernel Attention combines multi-scale convolutions with selective attention to capture speaker information across temporal and frequency ranges. ERes2NetV2~\cite{chen2024eres2netv2} further improves multi-scale feature fusion with low computational cost, and SRC-TDNN~\cite{deng2026src} strengthens temporal modeling for short speech. Pre-trained speech models have also been combined with multi-resolution encoders to improve speaker modeling under limited speech conditions~\cite{myoung2026short}. Despite these advances, most existing methods still focus on pooled utterance-level embeddings, while direct modeling of frame-level speaker representations remains limited.

Existing speaker representation learning methods mainly optimize the utterance-level embedding space with discriminative objectives. Supervised approaches employ speaker classification losses, including AM-Softmax~\cite{wang2018additive}, AAM-Softmax~\cite{deng2019arcface}, SF-Softmax~\cite{han2023exploring}, and HAM-Softmax~\cite{fang2026hyperbolic}, while self-supervised methods learn from unlabeled speech through contrastive learning or self-distillation, such as SimCLR~\cite{zhang2021contrastive}, MoCo~\cite{xia2021self}, and DINO~\cite{cho2022non}. Knowledge distillation~\cite{li2026enhancing} and pre-trained speech model transfer have also been widely studied~\cite{vaessen2022fine, radford2023robust, zhao2024whisper, yin2026whisperdiari, han2025leveraging}. Nevertheless, these objectives mainly enforce speaker classification or embedding consistency, with limited consideration of fine-grained speaker variability and its disentanglement from linguistic content.

\subsection{Generative Representation Learning}
Generative representation learning models data distributions with latent variables and uses reconstruction objectives to constrain the latent space, enabling the learned representations to be used for audio and video generation~\cite{zhu2026omnihuman}. Variational Autoencoder (VAE)~\cite{kingma2013auto} is a representative approach. Given an input $\mathbf{x}$, the VAE encoder estimates the posterior distribution of the latent variable, as given in Eq.~\eqref{eq:vae_posterior}:
\begin{equation}
\label{eq:vae_posterior}
q_{\phi}(\mathbf{z}\mid\mathbf{x})
=
\mathcal{N}\left(
\boldsymbol{\mu}_{\phi}(\mathbf{x}),
\operatorname{diag}\left(
\boldsymbol{\sigma}_{\phi}^{2}(\mathbf{x})
\right)
\right),
\end{equation}
and samples the latent variable through reparameterization, as shown in Eq.~\eqref{eq:vae_reparameterization}:
\begin{equation}
\label{eq:vae_reparameterization}
\mathbf{z}
=
\boldsymbol{\mu}_{\phi}(\mathbf{x})
+
\boldsymbol{\sigma}_{\phi}(\mathbf{x})
\odot
\boldsymbol{\epsilon},
\qquad
\boldsymbol{\epsilon}
\sim
\mathcal{N}(\mathbf{0},\mathbf{I}).
\end{equation}

The decoder then reconstructs the input from $\mathbf{z}$. The VAE objective consists of a reconstruction term and a KL divergence term, as formulated in Eq.~\eqref{eq:vae_objective}:
\begin{equation}
\mathcal{L}_{\mathrm{VAE}}
=
\mathbb{E}_{q_{\phi}(\mathbf{z}\mid\mathbf{x})}
\left[
-\log p_{\theta}(\mathbf{x}\mid\mathbf{z})
\right]
+
\beta D_{\mathrm{KL}}
\left(
q_{\phi}(\mathbf{z}\mid\mathbf{x})
\parallel
p(\mathbf{z})
\right).
\label{eq:vae_objective}
\end{equation}

Based on this probabilistic latent framework, VAE and its variants have been widely explored for speech representation learning and factor disentanglement. FHVAE~\cite{hsu2017unsupervised} employs hierarchical latent variables to model speech variations at different temporal scales, learning structured and disentangled representations. ACVAE-VC~\cite{kameoka2019acvae} and related methods apply VAE-based latent modeling to non-parallel voice conversion and introduce auxiliary constraints to improve the separation of different speech factors. VQ-VAE~\cite{van2017neural} and its subsequent variants~\cite{wang2023neural} further employ discrete latent variables to model speech content and phonetic information in a structured manner. These studies demonstrate that latent-variable modeling provides an effective way to learn structured representations of different speech factors.

More recently, diffusion models~\cite{ho2020denoising} and flow matching~\cite{lipman2022flow} have emerged as effective generative approaches for modeling complex speech distributions. Diffusion models gradually add Gaussian noise to the original data and learn a reverse denoising process. Given clean data $\mathbf{x}_0$, the forward diffusion process can be expressed as Eq.~\eqref{eq:diffusion_forward}:
\begin{equation}
\label{eq:diffusion_forward}
q(\mathbf{x}_t \mid \mathbf{x}_0)
=
\mathcal{N}
\left(
\sqrt{\bar{\alpha}_t}\mathbf{x}_0,
(1-\bar{\alpha}_t)\mathbf{I}
\right),
\end{equation}
where $\mathbf{x}_t$ denotes the noisy representation at timestep $t$, and $\bar{\alpha}_t$ denotes the cumulative noise schedule coefficient. The reverse process estimates the denoising distribution as Eq.~\eqref{eq:diffusion_reverse}:
\begin{equation}
\label{eq:diffusion_reverse}
\boldsymbol{\mu}_{\theta}(\mathbf{x}_t,t)
=
\frac{1}{\sqrt{\alpha_t}}
\left(
\mathbf{x}_t
-
\frac{1-\alpha_t}
{\sqrt{1-\bar{\alpha}_t}}
\boldsymbol{\epsilon}_{\theta}(\mathbf{x}_t,t)
\right),
\end{equation}
where $\alpha_t$ denotes the noise schedule coefficient at timestep $t$, and $\boldsymbol{\epsilon}_{\theta}(\mathbf{x}_t,t)$ denotes the noise predicted by the model.
In practice, diffusion models are commonly optimized using the noise prediction objective in Eq.~\eqref{eq:diffusion_loss}:
\begin{equation}
\label{eq:diffusion_loss}
\mathcal{L}_{\mathrm{diff}}
=
\mathbb{E}_{\mathbf{x}_0,t,\boldsymbol{\epsilon}}
\left[
\left\|
\boldsymbol{\epsilon}
-
\boldsymbol{\epsilon}_{\theta}(\mathbf{x}_t,t)
\right\|_2^2
\right],
\end{equation}
where $\boldsymbol{\epsilon}$ denotes the Gaussian noise added during the forward diffusion process.
Representative works such as DiffWave~\cite{kong2020diffwave} and
AudioLDM~\cite{liu2023audioldm} have demonstrated the effectiveness of
diffusion models for speech and audio generation.

Compared with diffusion models, flow matching directly learns a continuous vector field that transports samples from a simple prior distribution to the target data distribution. The corresponding continuous dynamics are described in Eq.~\eqref{eq:flow_ode}:
\begin{equation}
\label{eq:flow_ode}
\frac{d\mathbf{z}_t}{dt}
=
v_{\theta}(\mathbf{z}_t,t),
\end{equation}
where $\mathbf{z}_t$ denotes the intermediate state at time $t$, and $v_{\theta}(\mathbf{z}_t,t)$ denotes the learned vector field.
As formulated in Eq.~\eqref{eq:flow_matching}, the model is trained by matching the predicted vector field with a target vector field.
\begin{equation}
\label{eq:flow_matching}
\mathcal{L}_{\mathrm{FM}}
=
\mathbb{E}_{t,\mathbf{z}_t}
\left[
\left\|
v_{\theta}(\mathbf{z}_t,t)
-
u_t(\mathbf{z}_t)
\right\|_2^2
\right],
\end{equation}
where $u_t(\mathbf{z}_t)$ denotes the target vector field.
Recent flow-matching models, such as F5-TTS~\cite{chen2025f5}, have
achieved high-quality speech generation, demonstrating the potential of
flow matching for modeling complex speech distributions.

Generative modeling and representation disentanglement have recently been explored to learn more robust and discriminative speaker representations. Some studies~\cite{liu2023disentangling} disentangle voice and content information to improve the robustness of speaker representations. Although DSVAE~\cite{yadav2023dsvae} was developed for synthetic speech detection, it factorizes sequential latent variables into interpretable factors, including speaker-related information, using auxiliary constraints. Related latent diffusion methods further combine VAEs with diffusion models to represent different speech factors in the latent space~\cite{li2025disentangling}. These studies suggest that generative and disentangled modeling can help learn more robust speaker representations.

\section{Speaker Encoder with Local Temporal Modeling}
\begin{figure*}[!t]
    \centering
    \hspace*{-0.2cm}
    \includegraphics[width=17.5cm]{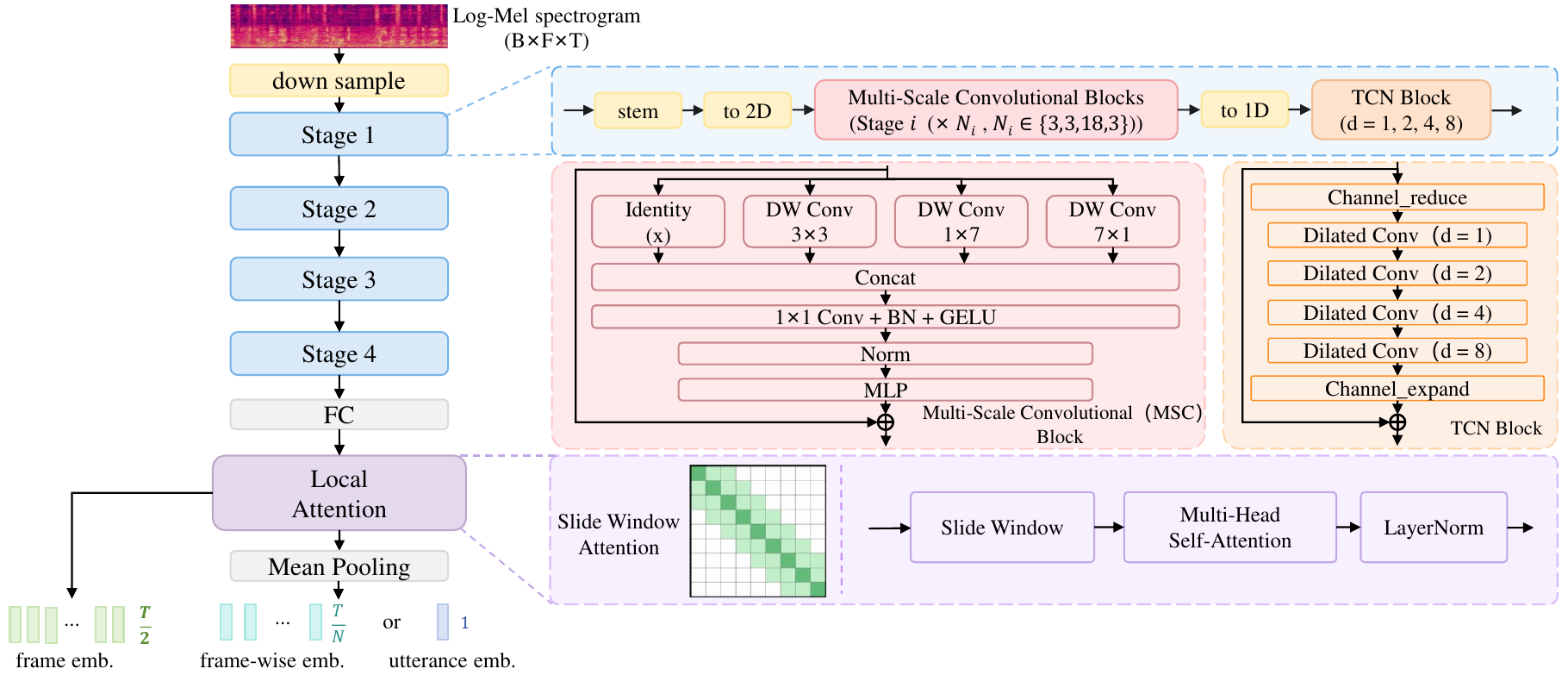}
    \caption{\textbf{Overview of the proposed speaker encoder architecture.} The input is downsampled by a factor of two and processed by a four-stage Multi-Scale Convolution (MSC) backbone with TCN-based local temporal modeling. The resulting features are projected to 192 dimensions and then processed by sliding-window self-attention to produce frame-level speaker representations with $T/2$ temporal steps. Mean pooling over local windows or the entire sequence produces frame-wise or utterance-level representations, respectively.}
    \label{fig:framework}
\end{figure*}

Limited local speaker evidence makes short-duration and frame-wise representations sensitive to phonetic and acoustic changes. Broader context provides additional evidence but may introduce cross-speaker interference. We therefore design a speaker encoder with hierarchical local temporal modeling (Fig.~\ref{fig:framework}) that combines a multi-scale convolutional backbone with TCN blocks~\cite{bai2018empirical} and Longformer-inspired sliding-window self-attention~\cite{beltagy2020longformer}. Short-range context improves local consistency and speaker discriminability while limiting cross-speaker mixing. Local mean pooling produces frame-wise speaker representations with adjustable temporal resolution, while global pooling yields an utterance-level embedding.

\subsection{Multi-Stage Encoder Backbone}
The proposed speaker encoder adopts a four-stage hierarchical feature extraction architecture. Given the input log-Mel filterbank features $\mathbf{X}_{\mathrm{in}}\in\mathbb{R}^{B\times 1 \times F\times T}$, the sequence is first downsampled along the time axis by a conv2d layer with stride 2, reducing its temporal length from $T$ to $T/2$. This process is formulated as Eq.~\eqref{eq:downsampling}
\begin{equation}
\label{eq:downsampling}
\mathbf{H}^{(0)}
=
F_{\mathrm{down}}(\mathbf{X}_{\mathrm{in}}),
\end{equation}
where $\mathbf{H}^{(0)} \in \mathbb{R}^{B \times C^{(0)} \times F \times T/2 }$ and $F_{\mathrm{down}}(\cdot)$ denotes the stride-2 downsampling layer and $D_0$ is the output feature dimension. The downsampled feature map is then reshaped into a one-dimensional temporal sequence $\mathbf{H}^{(0)}\in\mathbb{R}^{B\times D^{(0)}\times T/2}$, where $D^{(0)}=C^{(0)}F$. The resulting features are then processed by four encoder stages, each consisting of multiple Multi-Scale Convolutional(MSC) blocks and a TCN block. The MSC blocks $F_{\mathrm{MSC}}^{(i)}$ extract mlti-scale time-frequency features, while the TCN block  $F_{\mathrm{TCN}}^{(i)}$ models local temporal dependencies with a large receptive field. The $s$-th encoder stage is expressed as Eq.~\eqref{eq:encoder_stage}:
\begin{equation}
\begin{aligned}
\widetilde{\mathbf{H}}^{(i)}
&=
F_{\mathrm{MSC}}^{(i)}
\left(
F_{\mathrm{stem}}^{(i)}
\left(
F_{\mathrm{to2d}}^{(i)}
\left(
\mathbf{H}^{(i-1)}
\right)
\right)
\right), \\
\mathbf{H}^{(i)}
&=
F_{\mathrm{TCN}}^{(i)}
\left(
F_{\mathrm{to1d}}^{(i)}
\left(
\widetilde{\mathbf{H}}^{(i)}
\right)
\right),
\qquad i=1,\ldots,4,
\end{aligned}
\label{eq:encoder_stage}
\end{equation}
where $F_{\mathrm{stem}}^{(i)}(\cdot)$ denotes the 2D convolutional stem before the MSC blocks. The frequency strides of the four stages are set to $[1,2,2,2]$. When the frequency strdie is downsampled by a factor of 2, the channel dimension is doubled, keeping the product of the channel and frequency dimensions constant. $\widetilde{\mathbf{H}}^{(i)} \in\mathbb{R}^{B\times C^{(i)}\times F^{(i)}\times T/2}$ denotes the 2D feature map after the stem and MSC blocks, while
$\mathbf{H}^{(i)}
\in\mathbb{R}^{B\times D^{(i)}\times T/2}$,
where $D^{(i)}=C^{(i)}F^{(i)}$.
$F_{\mathrm{to2d}}^{(i)}(\cdot)$ and
$F_{\mathrm{to1d}}^{(i)}(\cdot)$ denote the feature reshaping operations.

\subsubsection{Multi-Scale Convolutional Block}
\leavevmode\par
Inspired by InceptionNeXt~\cite{yu2024inceptionnext}, we specially design a
Multi-Scale Convolutional (MSC) backbone by adapting its depthwise
convolution branches and feature-fusion scheme, as shown in
Fig.~\ref{fig:framework}. Different kernels capture local joint
time-frequency patterns, broader temporal and frequency context while maintaining low parameter and computational costs.

Given an input feature map
$\mathbf{H}_{\mathrm{MSC}}^{\mathrm{in}}
\in\mathbb{R}^{B\times C\times F\times T}$, the token mixer uses four
parallel branches: an identity mapping and $3\times3$, $1\times7$, and
$7\times1$ depthwise convolutions. The $3\times3$ branch captures local
time-frequency patterns, while the $1\times7$ and $7\times1$ branches
capture broader temporal and frequency context, respectively. Their
outputs are concatenated along the channel dimension to form a
$4C$-channel feature map, which is projected back to $C$ channels by a
$1\times1$ convolution followed by Batch Normalization and GELU
activation. This process is formulated as
Eq.~\eqref{eq:MSC_mixer}:
\begin{equation}
\begin{aligned}
\mathbf{H}_{\mathrm{MSC}}^{\mathrm{mix}}
&= \\[-3pt]
&\hspace{-2em}
F_{\mathrm{proj}}
\left(
\operatorname{Concat}
\left[
\mathbf{H}_{\mathrm{in}},
F_{3\times3}(\mathbf{H}_{\mathrm{in}}),
F_{1\times7}(\mathbf{H}_{\mathrm{in}}),
F_{7\times1}(\mathbf{H}_{\mathrm{in}})
\right]
\right),
\end{aligned}
\label{eq:MSC_mixer}
\end{equation}
where $\mathbf{H}_{\mathrm{MSC}}^{\mathrm{mix}}\in\mathbb{R}^{B\times C\times F\times T}$ denotes the fused feature map. Batch Normalization and a two-layer $1\times1$ convolutional MLP then transform this feature map, with a hidden dimension of $4C$ and GELU activation. The output is passed through DropPath and added to the input through a residual connection, as defined in Eq.~\eqref{eq:MSC_block}:
\begin{equation}
\label{eq:MSC_block}
\mathbf{H}_{\mathrm{MSC}}^{\mathrm{out}}
=
F_{\mathrm{drop}}
\left(
F_{\mathrm{MLP}}
\left(
F_{\mathrm{BN}}(\mathbf{H}_{\mathrm{MSC}}^{\mathrm{mix}})
\right)
\right)
+
\alpha_{\mathrm{MSC}}^{\mathrm{res}}\mathbf{H}_{\mathrm{MSC}}^{\mathrm{in}},
\end{equation}
where $\mathbf{H}_{\mathrm{Inc}}^{\mathrm{out}}\in\mathbb{R}^{B\times C\times F\times T}$ denotes the output feature map of the MSC block. Following the original model configuration and considering both parameter size and computational cost, the numbers of MSC Blocks are set to $[3,3,18,3]$.

\subsubsection{Local Temporal TCN Block}
\leavevmode\par
To model local temporal dependencies in the encoded features, we insert a TCN block after each encoder stage. Given an input feature map $\mathbf{H}_{\mathrm{TCN}}^{\mathrm{in}}\in\mathbb{R}^{B\times D\times T}$, the module adopts a bottleneck structure and first uses a $1\times1$
convolution to reduce the channel dimension before temporal modeling, which is formulated as Eq.~\eqref{eq:tcn_reduce}:
\begin{equation}
\label{eq:tcn_reduce}
\mathbf{H}_{\mathrm{TCN}}^{(0)}
=
F_{\mathrm{reduce}}
\left(
\mathbf{H}_{\mathrm{TCN}}^{\mathrm{in}}
\right),
\end{equation}
where $\mathbf{H}_{\mathrm{TCN}}^{(0)}\in\mathbb{R}^{B\times D_r\times T}$, and $D_r=D/8$.

The reduced features pass through four sequential TCN blocks. Each block
expands the channels from $D_r$ to $2D_r$, applies a depthwise dilated
convolution with kernel size 3, and projects the features back to
$D_r$ channels. A residual connection then adds the block input to its
output. The dilated convolution in the $l$-th block is defined as
Eq.~\eqref{eq:tcn_block}:
\begin{equation}
\label{eq:tcn_block}
\mathbf{H}_{\mathrm{TCN}}^{(i)}
=
F_{\mathrm{dil}}^{(i)}
\left(
\mathbf{H}_{\mathrm{TCN}}^{(i-1)}; d_i
\right),
\qquad
i=1,\ldots,4,
\end{equation}
where $F_{\mathrm{dil}}^{(l)}(\cdot)$ denotes the dilation convolution in the $l$-th TCN block, and $d_i$ denotes its dilation rate. The dilation rates are set to$[d_1,d_2,d_3,d_4]=[1,2,4,8]$, progressively expanding the temporal receptive field while preserving the output resolution.

After the four TCN blocks, a $1\times1$ convolution projects the features back to the original channel dimension, followed by a residual connection with the input, as defined in Eq.~\eqref{eq:tcn_output}:
\begin{equation}
\label{eq:tcn_output}
\mathbf{H}_{\mathrm{TCN}}^{\mathrm{out}}
=
F_{\mathrm{expand}}
\left(
\mathbf{H}_{\mathrm{TCN}}^{(4)}
\right)
+
\alpha_{\mathrm{TCN}}^{\mathrm{res}}\mathbf{H}_{\mathrm{TCN}}^{\mathrm{in}},
\end{equation}
where $\mathbf{H}_{\mathrm{TCN}}^{\mathrm{out}}\in\mathbb{R}^{B\times D\times T}$.

\begin{figure}[t]
    \centering
    \includegraphics[width=\columnwidth]{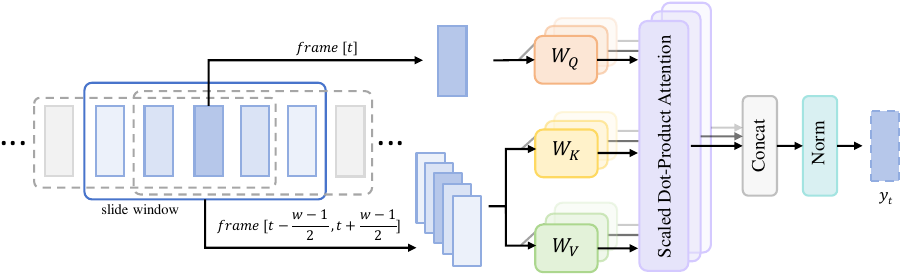}
    \caption{Illustration of the proposed local sliding-window
    self-attention module. Each frame attends only to the features within
    a local temporal window, which moves along the sequence with a
    stride of one.}
    \label{fig:local_attn}
\end{figure}

\subsubsection{Feature Reshaping}
\leavevmode\par
Inspired by the feature reshaping strategy in the ReDimNet, two reshaping operations, \textit{to1d} and \textit{to2d}, are used to convert features between 2D and 1D representations. Given a 2D feature map
$\widetilde{\mathbf{H}}\in\mathbb{R}^{B\times C\times F\times T}$,
the \textit{to1d} operation first permutes the feature dimensions and then merges the frequency and channel dimensions, producing
$\mathbf{H}\in\mathbb{R}^{B\times D\times T}$,
where $D=CF$.
Conversely, the \textit{to2d} operation reshapes the 1D representation back to
$\widetilde{\mathbf{H}}\in\mathbb{R}^{B\times C\times F\times T}$
and restores the corresponding dimension order.

\subsection{Slide Window Attention Modeling}
To enhance the stability of frame-level representations through local
dependencies among neighboring frames while reducing interference from
other speakers, we design a local sliding-window attention module, as
shown in Fig.~\ref{fig:local_attn}. The output of the convolutional
backbone is projected to obtain the module input $H_{\mathrm{loattn}}^{\mathrm{in}} \in\mathbb{R}^{B \times D_e \times T}$, where $B$ denotes the batch size, $D_e$ is the embedding dimension, and $T$ is the number of frames. In our implementation, $D_e=192$. The query, key, and value representations are obtained through $1\times1$ convolutional projections as Eq.~\eqref{eq:local_qkv}:
\begin{equation}
\mathbf{Q}=W_Q(\mathbf{H}_{\mathrm{loattn}}^{\mathrm{in}}),
\mathbf{K}=W_K(\mathbf{H}_{\mathrm{loattn}}^{\mathrm{in}}),
\mathbf{V}=W_V(\mathbf{H}_{\mathrm{loattn}}^{\mathrm{in}}),
\label{eq:local_qkv}
\end{equation}
where $\mathbf{Q},\mathbf{K},\mathbf{V}
\in \mathbb{R}^{B \times D_e \times T}$ denote the projected query, key, and value tensors, respectively. After transposing the temporal and feature dimensions, the query, key, and value vectors at frame $t$ are denoted by $\mathbf{q}_t,\mathbf{k}_t,\mathbf{v}_t \in \mathbb{R}^{B \times 1 \times D_e}$, respectively. The key and value matrices within the local window $\mathcal{W}_t$ are formed by stacking the corresponding frame-wise vectors, and are denoted by $\mathbf{K}_t,\mathbf{V}_t \in \mathbb{R}^{B \times W \times D_e}$, respectively.
Instead of performing attention over the entire sequence, each frame only
interacts with features within a local sliding window. For the $t$-th frame,
the local window is defined as Eq.~\eqref{eq:local_window}:
\begin{equation}
w_t
=
[t-P,\ldots,t,\ldots,t+P],
\qquad
P=\frac{w-1}{2},
\label{eq:local_window}
\end{equation}
where $w$ denotes the window size and the window slides with a stride of
1, so an attention output is computed for every input frame.
The key and value matrices within $w_t$ are denoted by
$\mathbf{K}_t,\mathbf{V}_t \in \mathbb{R}^{B \times w \times d}$,
respectively. The local attention output at frame $t$ is computed as Eq.~\eqref{eq:local_attention}:
\begin{equation}
\mathbf{y}_t
=
\operatorname{Softmax}
\left(
\frac{
\mathbf{q}_t
\mathbf{K}_t^{\mathrm{T}}
}{
\sqrt{d}
}
\right)
\mathbf{V}_t,
\label{eq:local_attention}
\end{equation}
where $d$ denotes the attention feature dimension. Invalid positions near
the sequence boundaries are masked before the softmax operation. The local attention outputs are finally normalized by Layer Normalization as Eq.~\eqref{eq:local_attention_output}:
\begin{equation}
\mathbf{H}_{\mathrm{local\_attn}}^{\mathrm{out}}
=
\operatorname{LN}(\mathbf{Y}),
\label{eq:local_attention_output}
\end{equation}
where
$\mathbf{Y}=[\mathbf{y}_1,\ldots,\mathbf{y}_T]$ and
$\mathbf{H}_{\mathrm{local\_attn}}^{\mathrm{out}}
\in\mathbb{R}^{B\times T\times D_e}$.
In our implementation, the window size is set to $W=51$.

\subsection{Multi-Resolution Speaker Representation}
Existing pooling methods, such as statistical pooling (e.g., TSTP~\cite{snyder2017deep}) and attention-based pooling, aggregate multi-frame information to obtain discriminative utterance-level embeddings. Their discrimination can rely on inter-frame variation, without requiring each frame embedding to be stable and speaker-discriminative. Mean pooling provides more stable frame-level features but may weaken utterance-level discrimination, while global temporal modeling can mix information from different speakers in multi-speaker speech. We therefore adopt local temporal modeling with sliding-window attention to capture nearby context while limiting cross-speaker interference, supporting mean pooling for frame-level, frame-wise, and utterance-level representation extraction

Given the output of the local attention module
$\mathbf{H}_{\mathrm{local\_attn}}^{\mathrm{out}}
\in\mathbb{R}^{B\times T\times D_e}$, each temporal feature
$\mathbf{h}_t\in\mathbb{R}^{D_e}$ is directly used as the frame-level
speaker representation. During training, the frame-level representations are aggregated by the temporal mean-pooling operator $P_{\mathrm{mean}}$ to obtain the full-utterance embedding $\mathbf{e}^{\mathrm{full}}=
P_{\mathrm{mean}}\big(\{\mathbf{h}_t\}_{t=1}^{T}\big)$, where $\mathbf{e}^{\mathrm{full}}\in\mathbb{R}^{B\times D_e}$.

During inference, frame-wise speaker representations can be extracted at
different temporal resolutions according to the downstream task. For every
$N$ consecutive temporal positions, mean pooling is applied as Eq.~\eqref{eq:framewise_rep}:
\begin{equation}
\mathbf{e}_{k}^{(N)}
=
\frac{1}{N}
\sum_{t=(k-1)N+1}^{kN}
\mathbf{h}_{t},
\qquad
k=1,\ldots,\left\lfloor\frac{T}{N}\right\rfloor,
\label{eq:framewise_rep}
\end{equation}
where $N$ determines the temporal resolution of the frame-wise representation.
If each temporal position corresponds to approximately $20\,\mathrm{ms}$,
the resulting representation covers approximately $20N\,\mathrm{ms}$ of speech.

\section{Generative Disentanglement Training Framework}
\begin{figure*}[!t]
    \centering
    \includegraphics[width=0.95\textwidth]{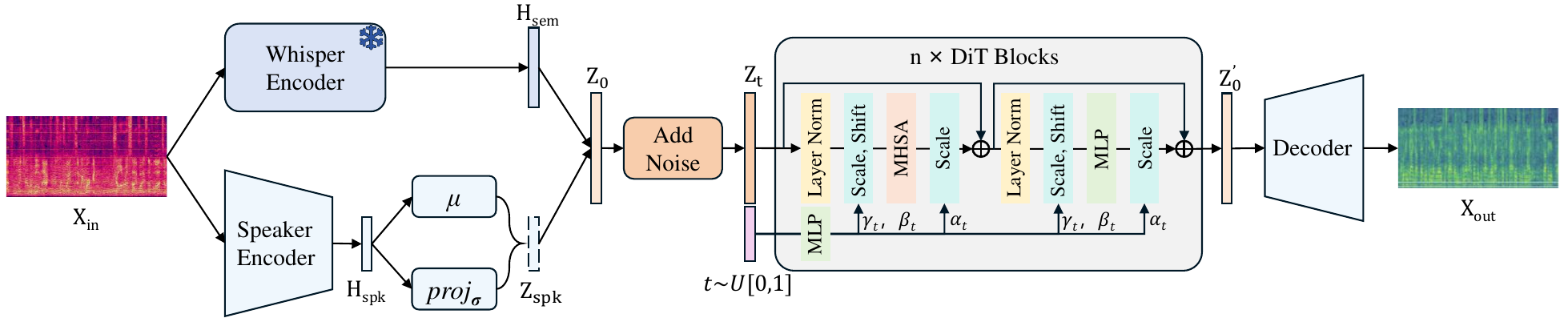}
    \caption{\textbf{Overview of the proposed generative disentanglement training framework.} Given an input log-Mel spectrogram, a frozen pretrained Whisper encoder extracts semantic features, while the speaker encoder extracts frame-level speaker features. A VAE-based module models the speaker representations and obtains latent speaker representations through reparameterization. After temporal alignment, both representations are concatenated to form the joint latent representation $\mathbf{Z}_0$. The noisy representation $\mathbf{Z}_t$ is then processed by DiT blocks to estimate $\mathbf{Z}'_0$, which is decoded to reconstruct the input.}
    \label{fig:train_framework}
\end{figure*}

Local temporal modeling strengthens speaker evidence, but frame-level features may still be sensitive to linguistic and acoustic variations. We therefore introduce the generative disentanglement framework in Fig.~\ref{fig:train_framework} to improve short-duration and frame-level speaker representations.

Given the input log-Mel spectrogram
$\mathbf{X}_{\mathrm{in}}\in\mathbb{R}^{B\times F\times T_{\mathrm{in}}}$,
we temporally divide it into two non-overlapping segments. The first half,
$\mathbf{X}_{\mathrm{sem}}$, is fed into a frozen pretrained whisper-tiny~\cite{radford2023robust} encoder to extract semantic representations
$\mathbf{H}_{\mathrm{sem}}\in\mathbb{R}^{B\times T\times D_{\mathrm{sem}}}$,
while the second half, $\mathbf{X}_{\mathrm{spk}}$, is processed by the
speaker encoder to obtain frame-level speaker representations
$\mathbf{H}_{\mathrm{spk}}\in\mathbb{R}^{B\times T\times D_{\mathrm{spk}}}$, which limits semantic leakage from Whisper and encourage the speaker encoder to retain complementary speaker-specific information.

A VAE-based module is introduced to model the latent distribution of the
frame-level speaker representations. It estimates
$\boldsymbol{\mu},\boldsymbol{\sigma}
\in\mathbb{R}^{B\times T\times D_{\mathrm{spk}}}$
and obtains the reparameterized representation
$\mathbf{H}'_{\mathrm{spk}}
\in\mathbb{R}^{B\times T\times D_{\mathrm{spk}}}$.
After temporal alignment, the semantic and speaker representations are
concatenated along the feature dimension to form
$\mathbf{Z}_0\in\mathbb{R}^{B\times T\times D_z}$,
where $D_z=D_{\mathrm{sem}}+D_{\mathrm{spk}}$.
The noisy latent representation
$\mathbf{Z}_t\in\mathbb{R}^{B\times T\times D_z}$
is then processed by $N$ DiT blocks to recover
$\mathbf{Z}'_0\in\mathbb{R}^{B\times T\times D_z}$,
which is decoded as $\mathbf{X}_{\mathrm{out}}\in\mathbb{R}^{B\times F\times T}$ to reconstruct
$\mathbf{X}_{\mathrm{sem}}\in\mathbb{R}^{B\times F\times T}$.

During training, the VAE regularizes the latent distribution, while flow matching learns a velocity field for generation. These objectives aim to characterize speaker-related variation and reduce non-speaker interference, improving representation stability and speaker discriminability.

\subsection{VAE-Based Speaker Representation Modeling}
To model the variation and uncertainty in short-duration and frame-level
speaker representations, we introduce a VAE-based formulation that maps
deterministic speaker features into a stochastic latent space. Given the
frame-level speaker sequence
$\mathbf{H}_{\mathrm{spk}}
=
[\mathbf{h}_{\mathrm{spk},1},\ldots,\mathbf{h}_{\mathrm{spk},T}]$,
the latent variable at the $i$-th temporal position is modeled by the
Gaussian approximate posterior in Eq.~\eqref{eq:spk_posterior}:
\begin{equation}
q_{\phi}
\left(
\mathbf{z}_{\mathrm{spk},i}
\mid
\mathbf{h}_{\mathrm{spk},i}
\right)
=
\mathcal{N}
\left(
\boldsymbol{\mu}_{i},
\operatorname{diag}
\left(
\boldsymbol{\sigma}_{i}^{2}
\right)
\right),
\label{eq:spk_posterior}
\end{equation}
where the frame-level speaker representation is used as the posterior mean,
$\boldsymbol{\mu}_{i}=\mathbf{h}_{\mathrm{spk},i}$, while
$\boldsymbol{\sigma}_{i}$ is estimated from the same representation.
Assuming conditional independence across temporal positions, the posterior
of the speaker latent sequence is factorized as
Eq.~\eqref{eq:spk_factorized_posterior}:
\begin{equation}
q_{\phi}
\left(
\mathbf{Z}_{\mathrm{spk}}
\mid
\mathbf{H}_{\mathrm{spk}}
\right)
=
\prod_{i=1}^{T}
q_{\phi}
\left(
\mathbf{z}_{\mathrm{spk},i}
\mid
\mathbf{h}_{\mathrm{spk},i}
\right),
\label{eq:spk_factorized_posterior}
\end{equation}
where $q_{\phi}(\mathbf{z}_{\mathrm{spk},i}\mid\mathbf{h}_{\mathrm{spk},i})$
denotes the approximate posterior of the speaker latent variable at the
$i$-th temporal position, while
$q_{\phi}(\mathbf{Z}_{\mathrm{spk}}\mid\mathbf{H}_{\mathrm{spk}})$
denotes the posterior over the complete speaker latent sequence.
To enable differentiable sampling, the reparameterization trick is applied as Eq.~\eqref{eq:spk_reparameterization}.
\begin{equation}
\mathbf{z}_{\mathrm{spk},i}
=
\boldsymbol{\mu}_{i}
+
\boldsymbol{\sigma}_{i}
\odot
\boldsymbol{\epsilon}_{i},
\qquad
\boldsymbol{\epsilon}_{i}
\sim
\mathcal{N}
\left(
\mathbf{0},
\mathbf{I}
\right).
\label{eq:spk_reparameterization}
\end{equation}

We model semantic and speaker information as two factorized components
in a joint latent space, defined as Eq.~\eqref{eq:joint_latent}:
\begin{equation}
\mathbf{Z}_{0}
\equiv
\left(
\mathbf{Z}_{\mathrm{sem}},
\mathbf{Z}_{\mathrm{spk}}
\right),
\label{eq:joint_latent}
\end{equation}
where
$\mathbf{Z}_{\mathrm{sem}}=\mathbf{H}_{\mathrm{sem}}$ is the semantic
representation extracted by a frozen Whisper-Tiny encoder, and
$\mathbf{Z}_{\mathrm{spk}}$ is the speaker latent representation
modeled by the VAE. After temporal alignment, the two components are
concatenated and jointly used by the generative model to reconstruct the
input.

Based on the definition of $\mathbf{Z}_{0}$, the joint distribution can
first be expanded as Eq.~\eqref{eq:joint_distribution}:
\begin{equation}
p_{\theta}
\left(
\mathbf{X}_{\mathrm{in}},
\mathbf{Z}_{0}
\right)
=
p_{\theta}
\left(
\mathbf{X}_{\mathrm{in}},
\mathbf{Z}_{\mathrm{sem}},
\mathbf{Z}_{\mathrm{spk}}
\right),
\label{eq:joint_distribution}
\end{equation}
since $\mathbf{Z}_{\mathrm{sem}}=\mathbf{H}_{\mathrm{sem}}$ is a fixed
deterministic representation, we do not assign it a variational
posterior or a separate prior. Instead, we condition the generative
model on $\mathbf{H}_{\mathrm{sem}}$ and define it as Eq.~\eqref{eq:joint_generation}:
\begin{equation}
p_{\theta}
\left(
\mathbf{X}_{\mathrm{in}},
\mathbf{Z}_{\mathrm{spk}}
\mid
\mathbf{H}_{\mathrm{sem}}
\right)
=
p
\left(
\mathbf{Z}_{\mathrm{spk}}
\right)
p_{\theta}
\left(
\mathbf{X}_{\mathrm{in}}
\mid
\mathbf{H}_{\mathrm{sem}},
\mathbf{Z}_{\mathrm{spk}}
\right),
\label{eq:joint_generation}
\end{equation}
where $p(\mathbf{Z}_{\mathrm{spk}})$ denotes the prior distribution of
the speaker latent representation, and
$p_{\theta}(\mathbf{X}_{\mathrm{in}}\mid
\mathbf{H}_{\mathrm{sem}},\mathbf{Z}_{\mathrm{spk}})$ models the
reconstruction of the input log-Mel spectrogram from the semantic and
speaker representations.

For completeness, we first consider a factorized variational formulation
for the joint latent representation. The approximate posterior is written as Eq.~\eqref{eq:joint_posterior}:
\begin{equation}
q
\left(
\mathbf{Z}_{0}
\mid
\mathbf{X}_{\mathrm{in}}
\right)
=
q_{\mathrm{sem}}
\left(
\mathbf{Z}_{\mathrm{sem}}
\mid
\mathbf{X}_{\mathrm{in}}
\right)
q_{\phi}
\left(
\mathbf{Z}_{\mathrm{spk}}
\mid
\mathbf{H}_{\mathrm{spk}}
\right),
\label{eq:joint_posterior}
\end{equation}
and the corresponding evidence lower bound is Eq.~\eqref{eq:joint_elbo}:
\begin{equation}
\begin{aligned}
\log p_{\theta}
\left(
\mathbf{X}_{\mathrm{in}}
\right)
\geq\;&
\mathbb{E}_{
q(\mathbf{Z}_{0}\mid\mathbf{X}_{\mathrm{in}})
}
\left[
\log p_{\theta}
\left(
\mathbf{X}_{\mathrm{in}}
\mid
\mathbf{Z}_{\mathrm{sem}},
\mathbf{Z}_{\mathrm{spk}}
\right)
\right]
\\
&-
D_{\mathrm{KL}}
\left[
q_{\mathrm{sem}}
\left(
\mathbf{Z}_{\mathrm{sem}}
\mid
\mathbf{X}_{\mathrm{in}}
\right)
\middle\|
p
\left(
\mathbf{Z}_{\mathrm{sem}}
\right)
\right]
\\
&-
D_{\mathrm{KL}}
\left[
q_{\phi}
\left(
\mathbf{Z}_{\mathrm{spk}}
\mid
\mathbf{H}_{\mathrm{spk}}
\right)
\middle\|
p
\left(
\mathbf{Z}_{\mathrm{spk}}
\right)
\right].
\end{aligned}
\label{eq:joint_elbo}
\end{equation}

The first term is the reconstruction objective, while the KL terms
regularize the latent distributions. In practice,
$\mathbf{Z}_{\mathrm{sem}}=\mathbf{H}_{\mathrm{sem}}$ is a fixed
deterministic representation extracted by the frozen Whisper-Tiny
encoder. We therefore omit the semantic KL term and use the conditional
ELBO in Eq.~\eqref{eq:conditional_elbo}, which regularizes only the
speaker latent distribution.
\begin{equation}
\begin{aligned}
\log p_{\theta}
\left(
\mathbf{X}_{\mathrm{in}}
\right)
\geq\;&
\mathbb{E}_{
q(\mathbf{Z}_{0}\mid\mathbf{X}_{\mathrm{in}})
}
\left[
\log p_{\theta}
\left(
\mathbf{X}_{\mathrm{in}}
\mid
\mathbf{H}_{\mathrm{sem}},
\mathbf{Z}_{\mathrm{spk}}
\right)
\right]
\\
&-
D_{\mathrm{KL}}
\left[
q_{\phi}
\left(
\mathbf{Z}_{\mathrm{spk}}
\mid
\mathbf{H}_{\mathrm{spk}}
\right)
\middle\|
p
\left(
\mathbf{Z}_{\mathrm{spk}}
\right)
\right].
\end{aligned}
\label{eq:conditional_elbo}
\end{equation}

For each frame-level speaker latent variable, we use a Gaussian prior
centered at the speaker-class center:
$p(\mathbf{z}_{\mathrm{spk},i})
=\mathcal{N}(\mathbf{c}_{y},\mathbf{I})$,
where $\mathbf{c}_{y}$ is the center of speaker class $y$. We omit the
speaker identity from the prior notation for simplicity. Assuming
factorized posterior and prior distributions across temporal positions,
the negative conditional ELBO gives the weighted VAE objective in
Eq.~\eqref{eq:vae_objective2}:
\begin{equation}
\begin{aligned}
\mathcal{L}_{\mathrm{VAE}}
=
&-
\mathbb{E}_{
q_{\phi}
\left(
\mathbf{Z}_{\mathrm{spk}}
\mid
\mathbf{H}_{\mathrm{spk}}
\right)
}
\left[
\log p_{\theta}
\left(
\mathbf{X}_{\mathrm{in}}
\mid
\mathbf{H}_{\mathrm{sem}},
\mathbf{Z}_{\mathrm{spk}}
\right)
\right]
\\
&+
\beta_{\mathrm{KL}}
\sum_{i=1}^{T}
D_{\mathrm{KL}}
\left[
q_{\phi}
\left(
\mathbf{z}_{\mathrm{spk},i}
\mid
\mathbf{h}_{\mathrm{spk},i}
\right)
\middle\|
p
\left(
\mathbf{z}_{\mathrm{spk},i}
\right)
\right],
\end{aligned}
\label{eq:vae_objective2}
\end{equation}
where $\beta_{\mathrm{KL}}$ controls the strength of the KL
regularization. The reconstruction term jointly uses semantic and
speaker information, while the KL term regularizes the frame-level
speaker latent distribution.

\subsection{Disentangled Generation with Flow Matching}
After temporal alignment, the stochastic speaker representation
$\mathbf{Z}_{\mathrm{spk}}$ is concatenated with the semantic
representation $\mathbf{H}_{\mathrm{sem}}$ extracted by the frozen
Whisper-Tiny encoder to form the joint latent representation
$\mathbf{Z}_{0}=\operatorname{Concat}
(\mathbf{H}_{\mathrm{sem}},\mathbf{Z}_{\mathrm{spk}})$.
Here, $\mathbf{H}_{\mathrm{sem}}$ encodes speech content, while
$\mathbf{Z}_{\mathrm{spk}}$ captures time-varying fine-grained speaker
information through VAE-based stochastic modeling.

To model the continuous generation process, we sample
$\boldsymbol{\epsilon}\sim\mathcal{N}(\mathbf{0},\mathbf{I})$ from a
standard Gaussian distribution and sample
$t\sim\mathcal{U}(0,1)$ as a continuous time variable. Given the clean
joint latent representation $\mathbf{Z}_{0}$, the noisy representation
at time $t$ is defined by the linear probability path as Eq.~\eqref{eq:linear_probability_path}:
\begin{equation}
\mathbf{Z}_{t}
=
(1-t)\mathbf{Z}_{0}
+
t\boldsymbol{\epsilon},
\qquad
t\in[0,1],
\label{eq:linear_probability_path}
\end{equation}
where $\mathbf{Z}_{t}$ denotes the interpolated latent representation.

At $t=0$, $\mathbf{Z}_{t}=\mathbf{Z}_{0}$ represents the clean
semantic--speaker representation, while at $t=1$,
$\mathbf{Z}_{t}=\boldsymbol{\epsilon}$ is standard Gaussian noise.
Differentiating the linear path gives the target velocity as Eq.~\eqref{eq:target_velocity_field}:
\begin{equation}
\mathbf{u}_{t}^{*}
=
\frac{\mathrm{d}\mathbf{Z}_{t}}{\mathrm{d}t}
=
\boldsymbol{\epsilon}-\mathbf{Z}_{0},
\label{eq:target_velocity_field}
\end{equation}
where $\mathbf{u}_{t}^{*}$ denotes the target velocity at time $t$. A
DiT-based~\cite{peebles2023scalable} velocity network
$\mathbf{v}_{\psi}(\mathbf{Z}_{t},t)$ predicts this velocity, and is
trained using the Flow Matching loss as Eq.~\eqref{eq:flow_matching_loss}:
\begin{equation}
\mathcal{L}_{\mathrm{FM}}
=
\mathbb{E}_{\mathbf{Z}_{0},\boldsymbol{\epsilon},t}
\left[
\left\|
\mathbf{v}_{\psi}(\mathbf{Z}_{t},t)
-
\left(
\boldsymbol{\epsilon}-\mathbf{Z}_{0}
\right)
\right\|_{2}^{2}
\right].
\label{eq:flow_matching_loss}
\end{equation}

Minimizing $\mathcal{L}_{\mathrm{FM}}$ trains the DiT to follow the
joint semantic--speaker latent path. The Flow Matching module and Mel
decoder are used only during training; at inference, the speaker
encoder and VAE directly extract fine-grained frame-level speaker
representations.

\subsection{Training Objectives}

We adopt a two-stage training strategy. In the first stage, the speaker
encoder is trained with AAM-Softmax using the full-utterance embedding
$\mathbf{e}^{\mathrm{full}}$. The corresponding loss is Eq.~\eqref{eq:aam_softmax_loss}:

\begin{equation}
\begin{array}{l}
\mathcal{L}_{\mathrm{AAM}} = \\[-2pt]
-\dfrac{1}{N}
\displaystyle\sum_{n=1}^{N}
\log
\dfrac{
\exp\left[s\cos\left(\theta_{y_n}+m\right)\right]
}{
\exp\left[s\cos\left(\theta_{y_n}+m\right)\right]
+
\displaystyle\sum_{j\neq y_n}
\exp\left[s\cos\left(\theta_j\right)\right]
},
\end{array}
\label{eq:aam_softmax_loss}
\end{equation}
where $N$ is the batch size, $y_n$ is the speaker label of the
$n$-th sample, and $\theta_{y_n}$ and $\theta_j$ are the angles between
the $\ell_2$-normalized $\mathbf{e}^{\mathrm{full}}$ and the target and
non-target class centers, respectively. Here, $m$ denotes the additive
angular margin, and $s$ scales the logits before the softmax operation.

In the second stage, the AAM-Softmax, VAE, and Flow Matching losses are jointly optimized to refine the frame-level speaker representations as Eq.~\eqref{eq:stage2_objective}:
\begin{equation}
\mathcal{L}_{\mathrm{stage2}}
=
\mathcal{L}_{\mathrm{AAM}}
+
\lambda\left(
\mathcal{L}_{\mathrm{VAE}}
+
\mathcal{L}_{\mathrm{FM}}
\right),
\label{eq:stage2_objective}
\end{equation}
where $\mathcal{L}_{\mathrm{AAM}}$ enforces speaker discrimination,
while $\mathcal{L}_{\mathrm{VAE}}$ and $\mathcal{L}_{\mathrm{FM}}$
jointly model and regularize the frame-level speaker latent space. The
coefficient $\lambda$ controls the overall contribution of these
generative losses relative to the AAM-Softmax loss. We set
$\lambda=1.0$ in all experiments.

\begin{table*}[t]
\centering
\caption{Full-utterance speaker verification results on three VoxCeleb1 test sets. \textbf{Bold} and \underline{underlined} values denote the best and second-best results.}
\label{tab:voxceleb1_full}

\footnotesize
\setlength{\tabcolsep}{6pt}
\renewcommand{\arraystretch}{0.9}

\begin{tabular}{lcc*{6}{c}}
\toprule
\multirow{2}{*}{\textbf{Model}}
&
\multirow{2}{*}{\textbf{Params}}
&
\multirow{2}{*}{\textbf{FLOPs}}
&
\multicolumn{2}{c}{\textbf{VoxCeleb1-O}}
&
\multicolumn{2}{c}{\textbf{VoxCeleb1-E}}
&
\multicolumn{2}{c}{\textbf{VoxCeleb1-H}}
\\
\cmidrule(lr){4-5}
\cmidrule(lr){6-7}
\cmidrule(lr){8-9}
& & &
\textbf{EER$\downarrow$}
& \textbf{minDCF$\downarrow$}
& \textbf{EER$\downarrow$}
& \textbf{minDCF$\downarrow$}
& \textbf{EER$\downarrow$}
& \textbf{minDCF$\downarrow$}
\\
\midrule
ECAPA-TDNN  & \underline{6.2M} & \textbf{1.0G} & 0.9839 & 0.0872 & 1.1169 & 0.1254 & 2.1191 & 0.2099 \\
CAM++       & 7.0M & \underline{1.1G} & 0.8350 & 0.1046 & 0.8595 & 0.0990 & 1.7672 & 0.1810 \\
ResNet      & 6.3M & 4.5G & 0.9995 & 0.1125 & 1.1141 & 0.1227 & 1.9872 & 0.1935 \\
Res2Net     & \textbf{4.0M} & 1.8G & 1.5424 & 0.1553 & 1.4521 & 0.1528 & 2.5355 & 0.2416 \\
ERes2Net    & 6.6M & 3.4G & 1.0956 & 0.1183 & 1.0906 & 0.1230 & 2.0665 & 0.2029 \\
ERes2NetV2  & 17.8M& 8.4G & 0.8188 & 0.0750 & 0.9120 & 0.1037 & 1.7189 & 0.1718 \\
ReDimNet-B4 & 6.4M & 7.2G & \textbf{0.5744} & \textbf{0.0548} & \textbf{0.8006} & \textbf{0.0891} & \textbf{1.4571} & \textbf{0.1408} \\
\midrule
Ours & 9.7M & 3.1G & \underline{0.6699} & \underline{0.0730} & \underline{0.8561} & \underline{0.0973} & \underline{1.6336} & \underline{0.1658}
\\
\bottomrule
\end{tabular}
\end{table*}

\begin{table*}[t]
\centering
\caption{Short-duration speaker verification results on the three VoxCeleb1 test sets. \textbf{Bold} and \underline{underlined} values denote the best and second-best results.}
\label{tab:voxceleb1_short}
\setlength{\tabcolsep}{0.7pt}
\renewcommand{\arraystretch}{1.00}
\resizebox{\textwidth}{!}{
\begin{tabular}{l*{18}{c}}
\toprule

\multirow{3}{*}{\textbf{Model}}
&
\multicolumn{6}{c}{\textbf{VoxCeleb1-O}}
&
\multicolumn{6}{c}{\textbf{VoxCeleb1-E}}
&
\multicolumn{6}{c}{\textbf{VoxCeleb1-H}}
\\
\cmidrule(lr){2-7}
\cmidrule(lr){8-13}
\cmidrule(lr){14-19}

&
\multicolumn{2}{c}{\textbf{0.5 s}}
&
\multicolumn{2}{c}{\textbf{1 s}}
&
\multicolumn{2}{c}{\textbf{2 s}}
&
\multicolumn{2}{c}{\textbf{0.5 s}}
&
\multicolumn{2}{c}{\textbf{1 s}}
&
\multicolumn{2}{c}{\textbf{2 s}}
&
\multicolumn{2}{c}{\textbf{0.5 s}}
&
\multicolumn{2}{c}{\textbf{1 s}}
&
\multicolumn{2}{c}{\textbf{2 s}}
\\
\cmidrule(lr){2-3}
\cmidrule(lr){4-5}
\cmidrule(lr){6-7}
\cmidrule(lr){8-9}
\cmidrule(lr){10-11}
\cmidrule(lr){12-13}
\cmidrule(lr){14-15}
\cmidrule(lr){16-17}
\cmidrule(lr){18-19}

&
\textbf{EER$\downarrow$} & \textbf{minDCF$\downarrow$}
&
\textbf{EER$\downarrow$} & \textbf{minDCF$\downarrow$}
&
\textbf{EER$\downarrow$} & \textbf{minDCF$\downarrow$}
&
\textbf{EER$\downarrow$} & \textbf{minDCF$\downarrow$}
&
\textbf{EER$\downarrow$} & \textbf{minDCF$\downarrow$}
&
\textbf{EER$\downarrow$} & \textbf{minDCF$\downarrow$}
&
\textbf{EER$\downarrow$} & \textbf{minDCF$\downarrow$}
&
\textbf{EER$\downarrow$} & \textbf{minDCF$\downarrow$}
&
\textbf{EER$\downarrow$} & \textbf{minDCF$\downarrow$}
\\
\midrule

ECAPA-TDNN
& 18.6114 & 0.9999 & 7.4124 & 0.6280 & 2.8777 & 0.3123
& 18.6634 & 1.0000 & 7.5910 & 0.6113 & 2.9587 & 0.3053
& 21.4904 & 1.0000 & 10.4887 & 0.7595 & 4.8889 & 0.4546
\\
CAM++
& 15.6938 & 0.9998 & 5.8622 & 0.5601 & 2.2323 & 0.2780
& 15.5746 & 0.9998 & 5.8523 & 0.5215 & 2.1656 & 0.2428
& 18.0948 & 0.9999 & 8.3271 & 0.6756 & 3.8753 & 0.3852
\\
ResNet 
& 16.3902 & 1.0000 & 7.1910 & 0.7537 & 3.4817 & 0.4207
& 16.1213 & 1.0000 & 7.0106 & 0.7749 & 3.5052 & 0.4395
& 21.5845 & 1.0000 & 11.2249 & 0.9060 & 6.0413 & 0.6032
\\
Res2Net
& 13.8496 & 0.9999 & 6.6693 & 0.6066 & 3.2889 & 0.3571
& 13.9450 & 1.0000 & 6.6692 & 0.5768 & 3.1752 & 0.3261
& 17.2292 & 1.0000 & 9.4023 & 0.7072 & 5.0585 & 0.4545
\\
Eres2Net 
& 12.3707 & 0.9137 & 5.5114 & 0.5305 & 2.5626 & 0.2978
& 12.3692 & 0.8864 & 5.5691 & 0.5111 & 2.5381 & 0.2756
& 16.1895 & 0.9681 & 8.5712 & 0.6767 & 4.4244 & 0.4089
\\
Eres2NetV2 
& 11.3227 & \underline{0.8694} & 4.8697 & 0.4818 & 2.1276 & 0.2498
& 11.2824 & \underline{0.8239} & 4.8485 & 0.4597 & 2.1213 & 0.2371
& 14.8930 & \underline{0.9211} & 7.5432 & 0.6118 & 3.7556 & 0.3601
\\
ReDimNet-B4
& \underline{11.2516} & 1.0000 & \underline{4.1672} & \underline{0.4043} & \textbf{1.6215} & \textbf{0.1824}
& \underline{11.1899} & 1.0000 & \underline{4.0841} & \underline{0.4165} & \textbf{1.7033} & \textbf{0.1967}
& \underline{14.5487} & 1.0000 & \underline{6.4371} & \underline{0.5862} & \textbf{3.0055} & \textbf{0.2932}
\\

\midrule

\textbf{Ours}
& \textbf{9.1226} & \textbf{0.7936} & \textbf{3.8427} & \textbf{0.4036} & \underline{1.7345} & \underline{0.2054}
& \textbf{9.1255} & \textbf{0.7513} & \textbf{3.9244} & \textbf{0.4085} & \underline{1.8551} & \underline{0.2155}
& \textbf{12.6721} & \textbf{0.8765} & \textbf{6.3854} & \textbf{0.5624} & \underline{3.3185} & \underline{0.3310}
\\

\bottomrule
\end{tabular}
}
\end{table*}

\section{Experiment Setup}
\subsection{Datasets}
We evaluate the proposed method on speaker verification and speaker
diarization tasks. For speaker verification, models are trained on the
VoxCeleb2~\cite{chung2018voxceleb2} and CN-Celeb~\cite{fan2020cn}\footnote{\label{fn:dataset-license}%
The CN-Celeb dataset is licensed under Attribution-ShareAlike 4.0 International.
The AMI-SDM dataset is licensed under CC BY 4.0. The authoris confirm that all uses of the above resources are strictly for academic research purposes and not for any commercial application.}
training sets, and evaluated on the VoxCeleb1~\cite{nagrani2020voxceleb}
and CN-Celeb test sets. VoxCeleb1 and VoxCeleb2 mainly contain English
speech, whereas CN-Celeb is a Chinese speaker recognition corpus. The
Equal Error Rate (EER) and minimum Detection Cost Function (minDCF) are
used as evaluation metrics.

For speaker diarization, VoxConverse~\cite{chung2020spot} and the AMI-SDM test set~\cite{carletta2005ami}\footnotemark[\getrefnumber{fn:dataset-license}] are used as English evaluation sets, while the AliMeeting~\cite{yu2022m2met} and AISHELL-4~\cite{fu2021aishell} test sets are used as Chinese evaluation sets. VoxConverse contains real-world multi-speaker recordings, and AMI-SDM contains meeting speech recorded with a single distant microphone. AliMeeting and AISHELL-4 contain Chinese meeting speech with multiple speakers and overlapping speech. The Diarization Error Rate (DER) is used as the evaluation metric.
\subsection{Implementation Details}
All input utterances are cropped to $6$~s and converted into
80-dimensional log-Mel spectrograms. The utterance is split into two
non-overlapping $3$-s segments, which are used to extract the semantic
and speaker representations, respectively.

Our training framework is built on 3D-Speaker. The baseline models,
including ECAPA-TDNN~\cite{desplanques2020ecapa}, CAM++~\cite{wang2023cam++}, ResNet, Res2Net~\cite{zhou2021resnext}, ERes2Net~\cite{chen2023enhanced}, and ERes2NetV2~\cite{chen2024eres2netv2}, are implemented using their corresponding configurations without large-margin fine-tuning provided by 3D-Speaker~\cite{chen20253d}. ReDimNet~\cite{yakovlev2024reshape} is implemented based on the WeSpeaker~\cite{wang2024advancing}. Considering both the number of parameters and computational cost, we select ReDimNet-B4 as the ReDimNet baseline.

The model is trained in two stages using SGD with a batch size of 384.
The first stage trains the speaker encoder with AAM-Softmax for 45 epochs,
decaying the learning rate from $0.2$ to $1\times10^{-3}$. The second
stage jointly optimizes the AAM-Softmax, VAE, and Flow Matching losses
for 40 epochs, with the learning rate decayed from $1\times10^{-3}$ to
$1\times10^{-5}$. All loss weights are set to one, and experiments are
conducted on eight NVIDIA RTX 5090 GPUs.

At inference time, the generative modules are removed, and the speaker encoder directly produces the utterance- and frame-level speaker representations.

\section{Experiment Results}
\subsection{Speaker Verification Results}
We evaluate speaker verification on the English VoxCeleb and Chinese
CN-Celeb datasets at three temporal granularities: full-utterance,
short-duration, and frame-wise. For frame-wise evaluation, frame-level
features are extracted from 3-s segments and pooled over successive
groups of $n$ adjacent frames to obtain multiple frame-wise
representations, following each model's original pooling setting. For
short-duration evaluation, 16 short chunks per utterance are evaluated
independently to obtain stable estimates.

\begin{table*}[t]
\centering
\caption{Frame-wise speaker verification results on the VoxCeleb1 test sets at
different temporal granularities. Here, $f$ denotes one output frame;
$\dagger$ and $\ddagger$ indicate $8\times$ and $4\times$ the reference
frame rate, respectively. \textbf{Bold} and \underline{underline} values denote the
best and second-best results at each frame rate, while \textnormal{\textit{ITALIC}} values
denote the best result at each aligned temporal span.}
\label{tab:voxceleb1_framewise}
\setlength{\tabcolsep}{0.7pt}
\renewcommand{\arraystretch}{1.00}
\resizebox{\textwidth}{!}{
\begin{tabular}{l*{18}{c}}
\toprule

\multirow{3}{*}{\textbf{Model}}
&
\multicolumn{6}{c}{\textbf{VoxCeleb1-O}}
&
\multicolumn{6}{c}{\textbf{VoxCeleb1-E}}
&
\multicolumn{6}{c}{\textbf{VoxCeleb1-H}}
\\
\cmidrule(lr){2-7}
\cmidrule(lr){8-13}
\cmidrule(lr){14-19}

&
\multicolumn{2}{c}{\textbf{2$f$}}
&
\multicolumn{2}{c}{\textbf{10$f$}}
&
\multicolumn{2}{c}{\textbf{15$f$}}
&
\multicolumn{2}{c}{\textbf{2$f$}}
&
\multicolumn{2}{c}{\textbf{10$f$}}
&
\multicolumn{2}{c}{\textbf{15$f$}}
&
\multicolumn{2}{c}{\textbf{2$f$}}
&
\multicolumn{2}{c}{\textbf{10$f$}}
&
\multicolumn{2}{c}{\textbf{15$f$}}
\\
\cmidrule(lr){2-3}
\cmidrule(lr){4-5}
\cmidrule(lr){6-7}
\cmidrule(lr){8-9}
\cmidrule(lr){10-11}
\cmidrule(lr){12-13}
\cmidrule(lr){14-15}
\cmidrule(lr){16-17}
\cmidrule(lr){18-19}

&
\textbf{EER$\downarrow$} & \textbf{minDCF$\downarrow$}
&
\textbf{EER$\downarrow$} & \textbf{minDCF$\downarrow$}
&
\textbf{EER$\downarrow$} & \textbf{minDCF$\downarrow$}
&
\textbf{EER$\downarrow$} & \textbf{minDCF$\downarrow$}
&
\textbf{EER$\downarrow$} & \textbf{minDCF$\downarrow$}
&
\textbf{EER$\downarrow$} & \textbf{minDCF$\downarrow$}
&
\textbf{EER$\downarrow$} & \textbf{minDCF$\downarrow$}
&
\textbf{EER$\downarrow$} & \textbf{minDCF$\downarrow$}
&
\textbf{EER$\downarrow$} & \textbf{minDCF$\downarrow$}
\\
\midrule

ECAPA-TDNN
& 47.8875 & 1.0000 & 44.9959 & 1.0000 & 38.2574 & 1.0000
& 48.0757 & 1.0000 & 44.9593 & 1.0000 & 38.2033 & 1.0000
& 48.8396 & 1.0000 & 46.0504 & 1.0000 & 39.9533 & 1.0000
\\

CAM++
& 32.7798 & 0.9995 & 19.5082 & 0.9991 & 15.1904 & 0.9987
& 33.5502 & 0.9996 & 19.9441 & 0.9994 & 15.3838 & 0.9994
& 37.8243 & 0.9999 & 24.0232 & 0.9999 & 19.2660 & 0.9999
\\

ResNet
& 31.0855 & 1.0000 & 9.1386 & 0.8964 & 6.1886 & 0.6246
& 30.6526 & 1.0000 & 8.7790 & 0.8867 & 5.9352 & 0.6862
& 35.1405 & 1.0000 & 13.6103 & 0.9692 & 9.7669 &  0.8474
\\

Res2Net
& 27.0117 & 0.9998 & 8.7471 & 0.7260 & 6.0596 & 0.5728
& 26.8183 & 0.9999 & 8.8374 & 0.6999 & 5.9441 & 0.5384
& 30.5054 & 1.0000 & 11.8146 & 0.8073 & 8.5443 & 0.6681
\\

ERes2Net
& 19.5277 & 0.9995 & \underline{6.8579} & 0.6185 & 4.7298 & 0.4845
& 19.5988 & 0.9993 & \underline{6.9034} & 0.6018 & 4.7150 & 0.4570
& 24.3489 & 0.9997 & 10.3387 & 0.7669 & 7.5124 & 0.6187
\\

ERes2NetV2
& \underline{18.1028} & \underline{0.9934} & 6.9425 & \underline{0.5957} & \underline{4.4969} & \underline{0.4347}
& \underline{18.1337} & \underline{0.9918} & 6.9656 & \underline{0.5789} & \underline{4.5932} & \underline{0.4311}
& \underline{22.3599} & \underline{0.9994} & \underline{9.9080} & \underline{0.7356} & \underline{7.1203} & \underline{0.5993}
\\

ReDimNet-B4
& 24.8767 & 0.9993 & 18.0088 & 0.9992 & 14.8268 & 0.9992
& 25.3587 & 0.9995 & 18.3804 & 0.9995 & 15.1070 & 0.9995
& 30.3916 & 0.9997 & 22.6897 & 0.9996 & 19.0621 & 0.9997
\\

\midrule

Ours
& \textbf{4.7404} & \textbf{0.4783} & \textbf{4.3751} & \textbf{0.4510} & \textbf{4.0691} & \textbf{0.4351}
& \textbf{4.8961} & \textbf{0.4644} & \textbf{4.4875} & \textbf{0.4354} & \textbf{4.2482} & \textbf{0.4220}
& \textbf{7.5270} & \textbf{0.6216} & \textbf{7.0265} & \textbf{0.5894} & \textbf{6.7156} & \textbf{0.5816}
\\
\midrule

\multicolumn{19}{c}{
\textit{Aligned temporal granularity}
\quad
($^{\dagger}\!:{f}\times 8$; $^{\ddagger}\!:{f}\times 4$)
}
\\
\midrule

ECAPA-TDNN$^{\dagger}$
& 36.3676 & 1.0000 & 10.2132 & 0.7639 & 6.4807 & 0.5743
& 36.1920 & 1.0000 & 10.3476 & 0.7363 & 6.5545 & 0.5617
& 38.0465 & 1.0000 & 13.0570 & 0.8551 & 9.0813 & 0.7067
\\
CAM++$^{\ddagger}$
& 22.0621 & 0.9991 & 6.4632 & 0.5910 & 4.2639 & 0.4693
& 22.5659 & 0.9995 & 6.6547 & 0.5800 & 4.3490 & 0.4305
& 26.6828 & 0.9998 & 9.7736 & 0.7525 & 6.8694 & 0.6124
\\
ReDimNet-B4$^{\ddagger}$
& 19.7184 & 0.9989 & 7.8930 & 0.7670 & 5.5198 & 0.5246
& 20.0387 & 0.9995 & 7.9324 & 0.7500 & 5.5462 & 0.5227
& 24.5603 & 0.9996 & 11.1018 & 0.8922 & 8.2934 & 0.6730
\\

\midrule

Ours$^{\ddagger}$
& \textit{4.4355} & \textit{0.4732} & \textit{3.0499} & \textit{0.3504} & \textit{2.3698} & \textit{0.2974}
& \textit{4.5560} & \textit{0.4424} & \textit{3.1833} & \textit{0.3459} & \textit{2.5876} & \textit{0.2868}
& \textit{7.1406} & \textit{0.5998} & \textit{5.3153} & \textit{0.4954} & \textit{4.4691} & \textit{0.4335}
\\
\bottomrule
\end{tabular}
}
\end{table*}
\begin{table*}[t]
\centering
\caption{Speaker verification results on the CN-Celeb test set at different granularities. Here, $f$ denotes one output frame; notation and highlighting follow Tables~\ref{tab:voxceleb1_full} and~\ref{tab:voxceleb1_short}.}
\label{tab:cnceleb_multigranularity}
\setlength{\tabcolsep}{3.8pt}
\renewcommand{\arraystretch}{1.00}

\resizebox{\textwidth}{!}{
\begin{tabular}{l*{14}{c}}
\toprule

\multirow{3}{*}{\textbf{Model}}
&
\multicolumn{2}{c}{\textbf{Full utterance}}
&
\multicolumn{6}{c}{\textbf{Short duration}}
&
\multicolumn{6}{c}{\textbf{Frame-wise}}
\\

\cmidrule(lr){2-3}
\cmidrule(lr){4-9}
\cmidrule(lr){10-15}

&
\multicolumn{2}{c}{\textbf{All}}
&
\multicolumn{2}{c}{\textbf{0.5 s}}
&
\multicolumn{2}{c}{\textbf{1 s}}
&
\multicolumn{2}{c}{\textbf{2 s}}
&
\multicolumn{2}{c}{\textbf{2$f$}}
&
\multicolumn{2}{c}{\textbf{10$f$}}
&
\multicolumn{2}{c}{\textbf{15$f$}}
\\

\cmidrule(lr){2-3}
\cmidrule(lr){4-5}
\cmidrule(lr){6-7}
\cmidrule(lr){8-9}
\cmidrule(lr){10-11}
\cmidrule(lr){12-13}
\cmidrule(lr){14-15}

&
\textbf{EER$\downarrow$}
& \textbf{minDCF$\downarrow$}
&
\textbf{EER$\downarrow$}
& \textbf{minDCF$\downarrow$}
&
\textbf{EER$\downarrow$}
& \textbf{minDCF$\downarrow$}
&
\textbf{EER$\downarrow$}
& \textbf{minDCF$\downarrow$}
&
\textbf{EER$\downarrow$}
& \textbf{minDCF$\downarrow$}
&
\textbf{EER$\downarrow$}
& \textbf{minDCF$\downarrow$}
&
\textbf{EER$\downarrow$}
& \textbf{minDCF$\downarrow$}
\\

\midrule

ECAPA-TDNN
& 8.8175 & 0.4467
& 30.6153 & 1.0000
& 23.8785 & 0.9715
& 18.1484 & 0.7324
& 46.1216 & 1.0000
& 41.1612 & 1.0000
& 37.6522 & 1.0000
\\

CAM++
& 7.7105 & 0.4354
& 27.6204 & 1.0000
& 21.2458 & 0.8688
& 16.0324 & 0.6951
& 36.2373 & 0.9968
& 29.9153 & 0.9764
& 28.2092 & 0.9630
\\

ResNet
& 7.9020 & 0.4305
& 29.0041 & 1.0000
& 21.3901 & 0.9999
& 16.7583 & 0.8200
& 34.1524 & 1.0000
& 22.6106 & 0.9999
& 18.6879 & 0.9241
\\

Res2Net
& 8.5610 & 0.4760
& 25.7131 & 0.9962
& 21.0476 & 0.8403
& 16.6229 & 0.7020
& 31.0691 & 0.9988
& 22.2623 & 0.8690
& 19.7029 & 0.8401
\\

ERes2Net
& 8.3018 & 0.4220
& 25.0591 & 0.9903
& 20.2718 & 0.8150
& 16.2662 & 0.6860
& \underline{28.3778} & 0.9908
& 21.1114 & 0.8585
& \underline{18.2267} & 0.7988
\\

ERes2NetV2
& 7.8626 & \textbf{0.4013}
& 24.8900 & \underline{0.9603}
& 19.9500 & 0.7820
& 15.8399 & \underline{0.6612}
& 29.3957 & \underline{0.9793}
& \underline{20.0092} & \textbf{0.8113}
& 18.3751 & \textbf{0.7512}
\\

ReDimNet-B4
& \textbf{7.5286} & 0.4090
& \underline{23.1400} & 0.9428
& \textbf{17.5750} & \textbf{0.7466}
& \textbf{13.7651} & \textbf{0.6309}
& 30.1735 & 0.9985
& 27.1552 & 0.9769
& 25.4920 & 0.9773
\\

\midrule

Ours
& \underline{7.6373} & \underline{0.4085}
& \textbf{20.9793} & \textbf{0.9356}
& \underline{17.6389} & \underline{0.7780}
& \underline{14.1925} & 0.6651
& \textbf{18.6194} & \textbf{0.8417}
& \textbf{18.3078} & \underline{0.8143}
& \textbf{17.2118} & \underline{0.7961}
\\

\midrule

\multicolumn{15}{c}{
\textit{Aligned temporal granularity}
\quad
($^{\dagger}\!: f\times 8$;
 $^{\ddagger}\!: f\times 4$)
}
\\

\midrule

ECAPA-TDNN$^{\dagger}$
& \multicolumn{8}{c}{--}
& 37.4112 & 1.0000
& 24.4160 & 0.9844
& 21.0804 & 0.8636
\\

CAM++$^{\ddagger}$
& \multicolumn{8}{c}{--}
& 31.6421 & 0.9845
& 22.0707 & 0.8812
& 18.9584 & 0.8078
\\

ReDimNet-B4$^{\ddagger}$
& \multicolumn{8}{c}{--}
& 27.6176 & 0.9924
& 21.7886 & 0.8872
& 18.3396 & 0.7900
\\

\midrule

Ours$^{\ddagger}$
& \multicolumn{8}{c}{--}
& \textit{17.1702} & \textit{0.7898}
& \textit{15.9542} & \textit{0.7501}
& \textit{15.0909} & \textit{0.7045}
\\

\bottomrule
\end{tabular}
}
\end{table*}

\subsubsection{utterance-level results on voxceleb}
\leavevmode\par
Table~\ref{tab:voxceleb1_full} reports the full-utterance results,
together with the parameter counts and computational costs of the
compared models. ReDimNet achieves the lowest EER, while the proposed
method delivers comparable performance with less than half its FLOPs
and outperforms the remaining baselines, demonstrating an effective
trade-off between verification accuracy and computational efficiency.

Table~\ref{tab:voxceleb1_short} presents the results under short- and ultra-short duration conditions. For 1-s and 0.5-s speech segments, the proposed model achieves the best performance in terms of EER and minDCF, showing clear improvements over the compared methods. The proposed method preserves competitive full-utterance performance while achieving its largest gains under short-duration conditions.

\subsubsection{frame-wise results on voxceleb}
\leavevmode\par
The frame-wise verification results are reported in
Table~\ref{tab:voxceleb1_framewise}. To account for different
sampling rates, we evaluate the compared systems under matched frame-rate
and temporal-span. The proposed method consistently
outperforms all baselines in both settings and remains stable across
frame-wise granularities from $2f$ to $60f$, indicating its
frame-level representations preserve stable and discriminative speaker
information.

\subsubsection{utterance-level results on cnceleb}
\leavevmode\par
Table~\ref{tab:cnceleb_multigranularity} reports the multi-granularity
speaker verification results on CN-Celeb. The proposed method remains
highly competitive for full-utterance inputs, with only a negligible
performance gap from the best system. It performs particularly well at
$0.5$~s and achieves near-best results at $1$~s and $2$~s, with only
minor differences from the best results. These findings are consistent
with those on VoxCeleb, confirming strong long-duration performance and
robustness to short-duration speech.

\subsubsection{frame-wise results on cnceleb}
\leavevmode\par
The frame-wise results are also reported in
Table~\ref{tab:cnceleb_multigranularity}. The proposed method achieves
the best performance across all evaluated settings and remains stable
over granularities from $2f$ to $60f$, which indicates that the frame-level representations retain discriminative speaker information at different time resolutions. Together with the VoxCeleb results, these findings demonstrate the effectiveness and cross-lingual robustness of the proposed method for fine-grained speaker verification.

\begin{table*}[t]
\centering
\caption{Zero-shot speaker diarization results under different sets and temporal settings. Results are reported as diarization error rate (DER, \%), where $f$ denotes one output frame. Notation and highlighting conventions follow
Tables~\ref{tab:voxceleb1_full} and~\ref{tab:voxceleb1_short}.}
\label{tab:diarization_results}

\setlength{\tabcolsep}{2.3pt}
\renewcommand{\arraystretch}{1.00}

\resizebox{\textwidth}{!}{
\begin{tabular}{l*{20}{c}}
\toprule

\multirow{3}{*}{\textbf{Model}}
& \multicolumn{5}{c}{\textbf{VoxConverse}}
& \multicolumn{5}{c}{\textbf{AMI-SDM}}
& \multicolumn{5}{c}{\textbf{AliMeeting}}
& \multicolumn{5}{c}{\textbf{AISHELL-4}}
\\
\cmidrule(lr){2-6}
\cmidrule(lr){7-11}
\cmidrule(lr){12-16}
\cmidrule(lr){17-21}

& \multicolumn{2}{c}{\textbf{Short chunk}}
& \multicolumn{3}{c}{\textbf{Frame-wise}}
& \multicolumn{2}{c}{\textbf{Short chunk}}
& \multicolumn{3}{c}{\textbf{Frame-wise}}
& \multicolumn{2}{c}{\textbf{Short chunk}}
& \multicolumn{3}{c}{\textbf{Frame-wise}}
& \multicolumn{2}{c}{\textbf{Short chunk}}
& \multicolumn{3}{c}{\textbf{Frame-wise}}
\\
\cmidrule(lr){2-3}
\cmidrule(lr){4-6}
\cmidrule(lr){7-8}
\cmidrule(lr){9-11}
\cmidrule(lr){12-13}
\cmidrule(lr){14-16}
\cmidrule(lr){17-18}
\cmidrule(lr){19-21}

& \textbf{0.3\,s} & \textbf{0.5\,s}
& \textbf{2\,f} & \textbf{10\,f} & \textbf{15\,f}
& \textbf{0.3\,s} & \textbf{0.5\,s}
& \textbf{2\,f} & \textbf{10\,f} & \textbf{15\,f}
& \textbf{0.3\,s} & \textbf{0.5\,s}
& \textbf{2\,f} & \textbf{10\,f} & \textbf{15\,f}
& \textbf{0.3\,s} & \textbf{0.5\,s}
& \textbf{2\,f} & \textbf{10\,f} & \textbf{15\,f}
\\
\midrule

ECAPA-TDNN
& 53.701 & 30.779 & 82.650 & 56.683 & 48.682
& 78.928 & 57.360 & 93.295 & 82.024 & 77.147
& 48.343 & 40.624 & 77.680 & 56.810 & 47.341
& 71.530 & 53.461 & 85.482 & 71.912 & 64.440
\\

CAM++
& 51.316 & 22.245 & 60.884 & 46.914 & 31.848
& 82.707 & 48.914 & 79.634 & 68.572 & 58.350
& 47.024 & 36.898 & 62.687 & 51.765 & 44.930
& 66.431 & 51.941 & 73.387 & 66.403 & 52.069
\\

ResNet
& 49.286 & 29.446 & 52.346 & 17.926 & 15.370
& 80.843 & 65.663 & 80.672 & 52.442 & 41.825
& 46.012 & 32.456 & 48.174 & 25.031 & 25.454
& 61.119 & 42.292 & 69.856 & 20.332 & 14.581
\\

Res2Net
& 44.036 & 20.517 & 53.058 & 12.566 & 12.248
& 80.572 & 47.318 & 73.938 & 31.002 & 24.535
& 43.965 & 29.698 & 45.084 & 21.702 & \underline{22.718}
& 56.534 & 30.120 & 58.630 & 15.961 & 11.620
\\

ERes2Net
& 37.116 & 19.613 & 41.373 & 12.315 & 11.955
& 67.602 & 42.048 & 69.038 & 28.369 & 23.999
& 42.738 & 28.701 & 43.073 & \underline{21.649} & \textit{\textbf{22.663}}
& \underline{51.827} & 29.679 & 59.282 & 12.729 & 10.548
\\

ERes2NetV2
& 33.002 & 16.436 & \underline{32.578} & \underline{11.343} & \textit{\underline{11.457}}
& 66.895 & 40.490 & \underline{56.329} & \underline{26.550} & \underline{24.154}
& \underline{40.317} & \textbf{27.507} & \underline{39.326} & \textit{\textbf{21.435}} & 22.917
& 55.484 & \textbf{27.648} & \underline{52.169} & \underline{12.127} & \underline{10.048}
\\

ReDimNet-B4
& \underline{30.136} & \textbf{15.742} & 63.695 & 60.090 & 56.783
& \underline{64.761} & \underline{37.266} & 81.766 & 78.917 & 73.696
& 42.732 & \underline{28.547} & 69.885 & 67.836 & 67.188
& 56.825 & \underline{27.857} & 74.599 & 66.802 & 61.716
\\

\midrule
Ours
& \textbf{27.054} & \underline{15.883} & \textbf{10.580} & \textbf{10.749} & \textbf{10.661}
& \textbf{52.754} & \textbf{35.485} & \textbf{22.255} & \textbf{22.050} & \textbf{22.107}
& \textbf{38.227} & 28.818 & \textbf{30.072} & 23.379 & 23.417
& \textbf{50.634} & 31.904 & \textbf{28.893} & \textbf{10.703} & \textit{\textbf{9.752}}
\\
\midrule

\multicolumn{21}{c}{
\textit{Aligned temporal granularity}
\quad
($^{\dagger}\!: f\times 8$;
 $^{\ddagger}\!: f\times 4$)
}
\\
\midrule

ECAPA-TDNN$^{\dagger}$
& \multicolumn{2}{c}{--} & 56.892 & 16.920 & 16.144
& \multicolumn{2}{c}{--} & 83.443 & 36.005 & 29.980
& \multicolumn{2}{c}{--} & 56.322 & 29.983 & 28.269
& \multicolumn{2}{c}{--} & 72.618 & 27.687 & 14.307
\\
CAM++$^{\ddagger}$
& \multicolumn{2}{c}{--} & 52.990 & 13.624 & 13.385
& \multicolumn{2}{c}{--} & 77.185 & 34.817 & 28.186
& \multicolumn{2}{c}{--} & 53.745 & 28.486 & 24.724
& \multicolumn{2}{c}{--} & 68.147 & 21.430 & 14.572
\\
ReDimNet-B4$^{\ddagger}$
& \multicolumn{2}{c}{--} & 60.446 & 48.802 & 47.113
& \multicolumn{2}{c}{--} & 80.599 & 71.885 & 69.751
& \multicolumn{2}{c}{--} & 68.657 & 65.665 & 64.355
& \multicolumn{2}{c}{--} & 68.660 & 50.055 & 45.948
\\

\midrule
Ours$^{\ddagger}$
& \multicolumn{2}{c}{--} & \textit{10.442} & \textit{10.996} & 12.729
& \multicolumn{2}{c}{--} & \textit{22.090} & \textit{22.358} & \textit{23.376}
& \multicolumn{2}{c}{--} & \textit{24.754} & 23.069 & 24.376
& \multicolumn{2}{c}{--} & \textit{12.849} & \textit{9.871}  & 10.480
\\
\bottomrule

\end{tabular}
}
\end{table*}

\subsection{Speaker Ddiarization Results}
We evaluate the proposed speaker representations on VoxConverse,
AMI-SDM, AliMeeting, and AISHELL-4 using a two-stage pipeline of speaker
representation extraction followed by spectral clustering. Models
trained on speaker recognition data are directly applied without
task-specific fine-tuning or additional diarization modules, yielding a
zero-shot evaluation. To reduce GPU memory usage, each recording is
processed in consecutive segments of approximately $60$~s.

For short-duration representations, we use $0.3$-s and $0.5$-s chunks
with strides of $0.15$~s and $0.25$~s, respectively. For frame-wise
representations, adjacent embeddings are pooled with non-overlapping
windows of $2$, $10$, and $15$ frames. Because the models have different
output frame rates, we report results for fixed frame counts and matched temporal spans. DER for all datasets and settings are in Table~\ref{tab:diarization_results}.

\begin{table}[t]
    \centering
    \caption{Ablation results on VoxCeleb1-O. EER (\%) is reported, \textbf{bold} indicates the best results; All denotes full-duration evaluation, $f$ the number of pooled frames, and $\star$ denotes an 80-ms per-frame span instead of 20 ms.}
    \label{tab:ablation_vox1o}
    \setlength{\tabcolsep}{2.2pt}
    \renewcommand{\arraystretch}{1.00}
    \resizebox{\columnwidth}{!}{
        \begin{tabular}{lccccccc}
            \toprule
            \multirow{2}{*}{Configuration}
            & \multicolumn{4}{c}{Utterance-level}
            & \multicolumn{3}{c}{Frame-level} \\
            \cmidrule(lr){2-5}
            \cmidrule(lr){6-8}
            & All
            & 0.5\,s
            & 1\,s
            & 2\,s
            & $2f$
            & $10f$
            & $15f$ \\
            \midrule
            ours
            & 0.6699 & 9.1226 & \textbf{3.8427} & \textbf{1.7345}  
            & 4.7404  & 4.3751  & \textbf{4.0691} \\
            w/o Generator
            & \textbf{0.6595} & 10.4581 & 4.3456 & 1.8426
            & 4.9340 & 4.6025 & 4.4305 \\
            Short-Utt. Training
            & 0.8935 & \textbf{8.7977} & 4.0441 & 2.0367 
            & \textbf{4.7039} & \textbf{4.3691} & 4.2554 \\
            Enc. + Frame Loss
            & 1.2441 & 8.9668 & 4.3400 & 2.3998
            & 7.0192 & 6.2570 & 6.0130 \\
            ERes2Netv2-based Enc.\textsuperscript{\(\star\)}
            & 0.7603 & 11.1286 & 5.0411 & 2.2295
            & 14.9580 & 5.6168 & 3.3659\\
            \bottomrule
        \end{tabular}
    }
\end{table}

\begin{table}[t]
    \centering
    \caption{sensitivity to the VAE KL weight $\beta$ on Vox1-O.
   EER (\%) is reported, \textbf{bold} indicates the best results.}
    \label{tab:beta-sensitivity}

    \setlength{\tabcolsep}{4.2pt}
    \renewcommand{\arraystretch}{1.00}

    \resizebox{\columnwidth}{!}{
        \begin{tabular}{cccccccc}
            \toprule
            \multirow[c]{2}{*}{VAE KL $(\beta_{\mathrm{KL}})$}
            & \multicolumn{4}{c}{Utterance-level}
            & \multicolumn{3}{c}{Frame-level} \\
            \cmidrule(lr){2-5}
            \cmidrule(lr){6-8}
            & All
            & 0.5\,s
            & 1\,s
            & 2\,s
            & $2f$
            & $10f$
            & $15f$ \\
            \midrule
            0.001
            & \textbf{0.6699} & 9.1226 & \textbf{3.8427} & \textbf{1.7345} & \textbf{4.7404} & \textbf{4.3751} & \textbf{4.0691} \\

            0.005
            & 0.7606 & \textbf{9.0343} & 4.0725 & 1.8223 & 5.0147 & 4.6926 & 4.4742 \\
        
            0.01
            & 0.6965 & 9.3228 & 4.0287 & 1.8163 & 4.8982 & 4.5469 & 4.3210 \\
        \bottomrule
    \end{tabular}
    }
\end{table}

\subsubsection{Results on English Datasets}
\leavevmode\par
For the English datasets, models trained on VoxCeleb2 are evaluated on
VoxConverse and AMI-SDM. The proposed method achieves competitive DERs
in short-duration settings and clear improvements in most frame-wise
settings. Its stable performance across temporal resolutions
demonstrates the discriminability and temporal robustness of the learned
frame-level speaker representations. On VoxConverse, the best
result outperforms the multi-stage 3D-Speaker system~\cite{chen20253d}, while its performance on AMI-SDM remains very close.

\subsubsection{Results on Chinese Datasets}
\leavevmode\par
For the Chinese datasets, the model is trained on CN-Celeb for speaker verification and evaluated on AliMeeting and AISHELL-4. On AliMeeting, the proposed method performs comparably to baselines such as ERes2NetV2, but remains stable in ultra-short and frame-wise settings. The limited gain may reflect the prevalence of long speaker turns, which leave limited room for further improvement. On AISHELL-4, the proposed method achieves the best overall DER, even better than 3D-Speaker system~\cite{chen20253d}, demonstrating its effectiveness across languages and recording conditions.

\subsection{Representation Visualization}
\begin{figure*}[t]
    \centering
    \includegraphics[
    width=\textwidth, 
    height=\textheight, 
    keepaspectratio
    ]{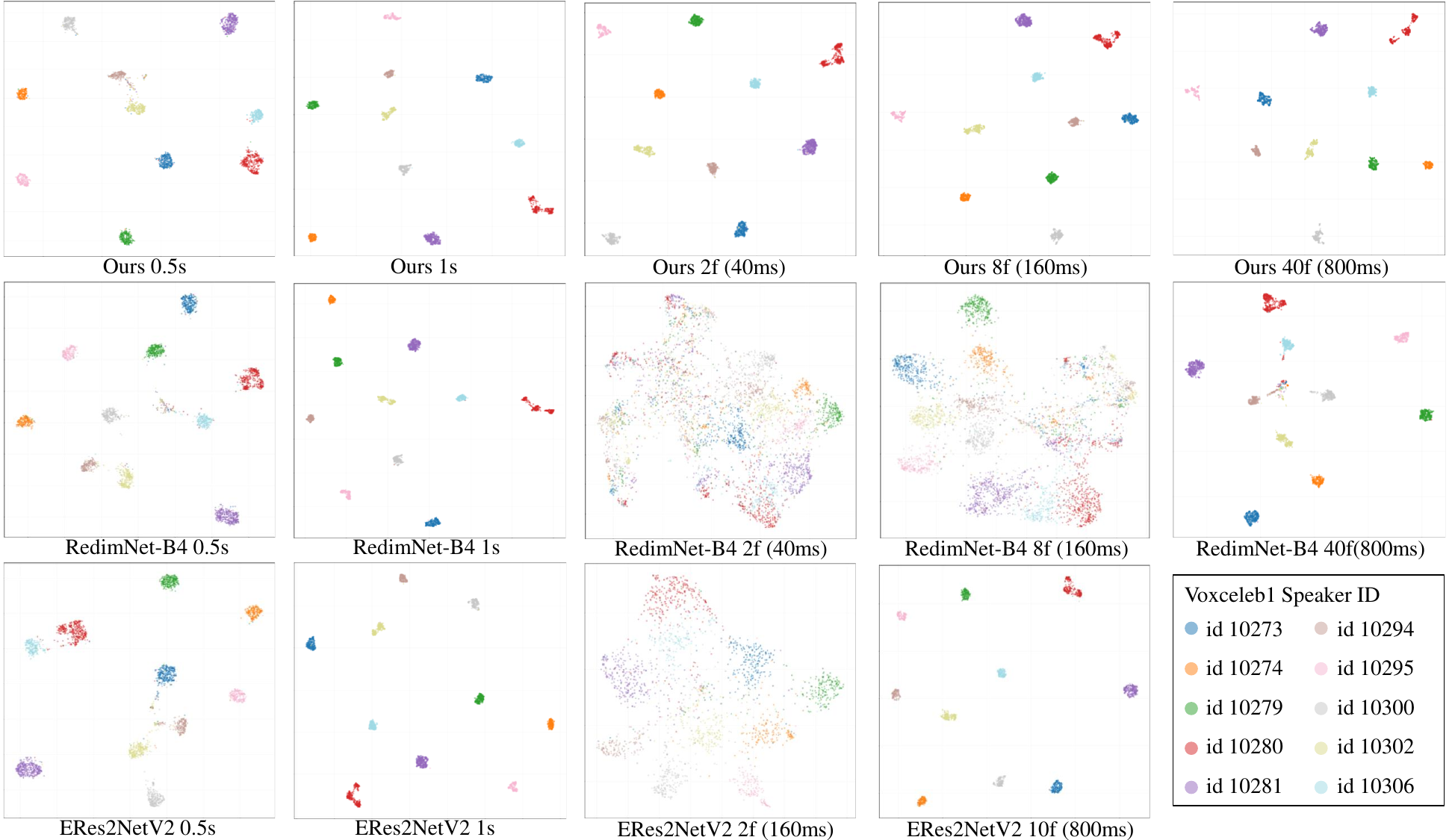}
    \caption{\textbf{UMAP visualization of speaker representations on VoxCeleb1.} Ten speakers are randomly selected, with 20 utterances from each speaker. Representations at different temporal granularities are projected into a two-dimensional space using UMAP. Different colors indicate different speakers, and $f$ denotes the number of output frames used for temporal pooling.}
    \label{fig:representation_visualization}
\end{figure*}

\begin{figure*}[t]
    \centering
    \includegraphics[
    width=\textwidth, 
    height=\textheight, 
    keepaspectratio
    ]{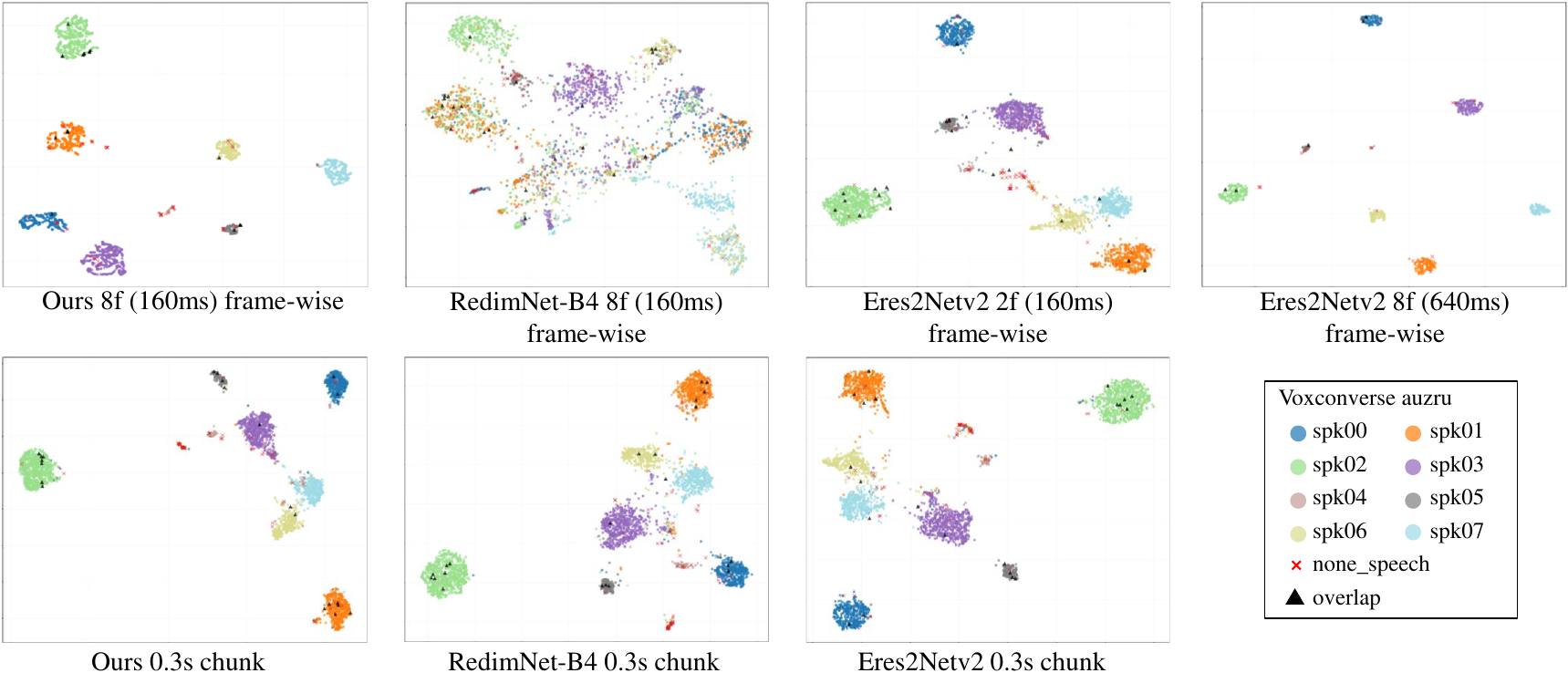}
    \caption{\textbf{UMAP visualization of speaker representations from the VoxConverse sample \textit{auzru}}. Representations at different temporal granularities are projected into a two-dimensional space. Colors denote different speakers, while red crosses and black triangles indicate non-speech and overlapping regions.}
    \label{fig:representation_visualization2}
\end{figure*}

\begin{table}[t]
    \centering
    \caption{Effect of the sliding-attention window size on Vox1-O. EER (\%) is reported, \textbf{bold} indicates the best results.}
    \label{tab:window-size-sensitivity}

    \setlength{\tabcolsep}{3.2pt}
    \renewcommand{\arraystretch}{1.00}

    \resizebox{\columnwidth}{!}{
        \begin{tabular}{cccccccc}
            \toprule
            \multirow[c]{2}{*}{slide window size}
            & \multicolumn{4}{c}{Utterance-level}
            & \multicolumn{3}{c}{Frame-level} \\
            \cmidrule(lr){2-5}
            \cmidrule(lr){6-8}
            & All
            & 0.5\,s
            & 1\,s
            & 2\,s
            & $2f$
            & $10f$
            & $15f$ \\
            \midrule
            25
            & 0.7872 & 10.6840 & 4.6295 & 2.0217 & 9.9363 & 8.7606 & 8.0910 \\
            35
            & 0.6646 & 10.5285 & 4.3782 & \textbf{1.7884} & 7.5289 & 6.7230 & 6.2751 \\
            45
            & 0.6593 & \textbf{10.4028} & 4.4231 & 1.8718 & 6.0834 & 5.5315 & 5.2874 \\
            51
            & 0.6595 & 10.4581 & \textbf{4.3456} & 1.8426 & \textbf{4.9340} & \textbf{4.6025} & 4.4305 \\
            55
            & \textbf{0.6446} & 10.5139 & 4.4477 & 1.9210 & 5.1163 & 4.6974 & \textbf{4.3724} \\
            \bottomrule
        \end{tabular}
    }
\end{table}

\subsubsection{Visualization for Speaker Verification}
\leavevmode\par
We use UMAP~\cite{mcinnes2018umap} to visualize speaker representations
on the VoxCeleb1 test set, comparing the proposed method with
ReDimNet-B4 and ERes2NetV2. The same 10 speakers and 20 utterances per
speaker are used for all models. For the $0.5$-s and $1$-s conditions,
16 temporally pooled representations are extracted from each utterance.
For frame-level visualization, all representations from a 3-s segment
are retained. Because the models have different temporal downsampling
rates, representations are compared at matched temporal spans of
approximately 40, 160, and 800 ms.

As shown in Fig.~\ref{fig:representation_visualization}, the proposed
method forms more compact within-speaker clusters and clearer
between-speaker separation than ReDimNet-B4 and ERes2NetV2, especially
under short-context conditions. For $0.5$-s inputs, it keeps
same-speaker samples closely grouped and provides clearer separation
between different speakers than the baselines. This advantage is more
pronounced for frame-level representations: clear speaker clusters are
maintained even at the 40-ms scale with only two output frames, and
remain stable at 160 ms and 800 ms. These results show that the proposed
encoder captures reliable speaker identity information from short
local contexts rather than depending mainly on long-range aggregation.

\begin{table}[t]
    \centering
    \caption{Sensitivity to the loss weight $\lambda$ on VoxCeleb1-O.
    EER (\%) is reported, \textbf{bold} indicates the best results.}
    \label{tab:lambda_sensitivity}

    \setlength{\tabcolsep}{4.2pt}
    \renewcommand{\arraystretch}{1.00}

    \resizebox{\columnwidth}{!}{
        \begin{tabular}{>{\centering\arraybackslash}m{0.85cm}ccccccc}
            \toprule
            \multirow[c]{2}{*}{$\lambda$}
            & \multicolumn{4}{c}{Utterance-level}
            & \multicolumn{3}{c}{Frame-level} \\
            \cmidrule(lr){2-5}
            \cmidrule(lr){6-8}
            & All
            & 0.5\,s
            & 1\,s
            & 2\,s
            & $2f$
            & $10f$
            & $15f$ \\
            \midrule
            0.8
            & \textbf{0.6380} & 9.4044 & 3.9876 & 1.7890 & 4.7933 & 4.4033 & 4.1756 \\

            1.0
            & 0.6699 & \textbf{9.1226} & \textbf{3.8427} & \textbf{1.7345} & 4.7404 & 4.3751  & 4.0691 \\

            1.2
            & 0.6858 & 9.6585 & 3.9888 & 1.8628 & \textbf{4.7088} & \textbf{4.0537} & \textbf{4.0537} \\
            \bottomrule
        \end{tabular}
    }
\end{table}

\subsubsection{Visualization for Speaker Diarization}
\leavevmode\par
We further visualize representations from the VoxConverse sample \textit{auzru}, which contains eight speakers, non-speech, and
overlapping speech. We compare $0.3$-s representations extracted with a
$0.15$-s hop and frame-level representations spanning approximately
160 ms. The recording is processed in consecutive $60$-s segments.
Because the models have different output frame rates, we also compare
eight-frame representations, which span 640 ms for ERes2NetV2 but only
160 ms for the proposed method.

As shown in Fig.~\ref{fig:representation_visualization2}, the proposed
method forms clear and compact speaker clusters for both the $0.3$-s
short-duration and 160-ms frame-level representations, despite
non-speech and overlapping speech. Compared with the baselines, it
shows stronger within-speaker compactness and between-speaker
separation. Notably, its eight-frame representation achieves separation
comparable to ERes2NetV2 while using only one quarter of the temporal
span (160 ms versus 640 ms). These results demonstrate that the proposed
frame-level representations retain speaker-discriminative information at
a much finer temporal resolution, supporting short-turn modeling and
precise speaker-change boundary localization.

\subsection{Ablation Study}
Table~\ref{tab:ablation_vox1o} presents the ablation results on
VoxCeleb1-O. We compare the complete proposed model with four variants:
(1) replacing the proposed encoder with ERes2Net, (2) removing the
generative module, (3) retaining only the encoder and adding a
frame-level speaker classification loss, and (4) training with
short-duration speech segments. All variants are evaluated under the
same speaker verification protocol, and performance is measured by EER.
The comparison shows that the complete model consistently achieves
better performance than its variants, confirming the effectiveness of
the proposed encoder and generative training framework.

Table~\ref{tab:beta-sensitivity} evaluates the VAE KL divergence weight
$\beta_{\mathrm{KL}}\in\{0.001,0.005,0.01\}$ on VoxCeleb1-O across utterance-level and frame-level conditions. Performance varies slightly, indicating low sensitivity to $\beta$. We use $\beta=0.001$ in subsequent experiments because it provides the best overall results while effectively regularizing the speaker latent distribution.

Table~\ref{tab:window-size-sensitivity} evaluates the sliding-window
attention size in the encoder-only training. Size of $w=51$
achieves the lowest EER under $2f$, $10f$ and 1-s conditions.
Considering its overall performance and the temporal resolution required
for speaker-change localization, we use $w=51$ in experiments.

Table~\ref{tab:lambda_sensitivity} evaluates the generative-loss weight
$\lambda$. Performance varies only slightly across the tested range,
with $\lambda=0.8$, $1.0$, and $1.2$ performing best for full-utterance,
short-duration, and frame-level conditions, respectively. Since no
single value is optimal for all settings, we set $\lambda=1.0$ for the
remaining experiments as a balanced choice.

\section{Conclusion}
In this work, we proposed FineSpeaker, a fine-grained speaker representation learning framework for speaker verification and speaker diarization. The proposed speaker encoder combines multi-scale feature extraction with local
temporal modeling to produce stable and discriminative utterance- and frame-level
speaker representations at flexible temporal resolutions. We further
introduced a generative disentanglement training scheme that combines a VAE
with Flow Matching. The VAE models the variation and uncertainty of the frame-level 
speaker representations, while Flow Matching regularizes their
distribution in a joint semantic--speaker latent space to obtain more stable
fine-grained representations. Speaker verification results on
VoxCeleb and CN-Celeb show that the proposed method maintains competitive
full-utterance performance and achieves clear improvements under
short-duration and frame-wise conditions. Zero-shot diarization
experiments on VoxConverse, AMI-SDM, AliMeeting, and AISHELL-4 further
demonstrate its effectiveness across languages, tasks, and temporal
granularities. Overall, the proposed framework provides stable,
discriminative, and generalizable fine-grained speaker representations
for short-duration speaker modeling and downstream diarization.

\bibliographystyle{IEEEtran}
\bibliography{reference}

@article{liu2023disentangling,
  title={Disentangling voice and content with self-supervision for speaker recognition},
  author={Liu, Tianchi and Lee, Kong Aik and Wang, Qiongqiong and Li, Haizhou},
  journal={Advances in Neural Information Processing Systems},
  volume={36},
  pages={50221--50236},
  year={2023}
}

@article{li2025disentangling,
  title={Disentangling speech representations learning with latent diffusion for speaker verification},
  author={Li, Zhe and Mak, Man-Wai and Chien, Jen-Tzung and Pilanci, Mert and Jin, Zezhong and Meng, Helen},
  journal={IEEE Transactions on Audio, Speech and Language Processing},
  year={2025},
  publisher={IEEE}
}

@article{wang2025review,
  title={A review of speaker verification: Methods, network architectures, tasks and challenges},
  author={Wang, Weijie and Zhao, Hong and Yang, Yikun and Yang, Yongjuan},
  journal={Engineering Applications of Artificial Intelligence},
  volume={162},
  pages={112351},
  year={2025},
  publisher={Elsevier}
}

@article{park2022review,
  title={A review of speaker diarization: Recent advances with deep learning},
  author={Park, Tae Jin and Kanda, Naoyuki and Dimitriadis, Dimitrios and Han, Kyu J and Watanabe, Shinji and Narayanan, Shrikanth},
  journal={Computer Speech \& Language},
  volume={72},
  pages={101317},
  year={2022},
  publisher={Elsevier}
}

@inproceedings{snyder2018x,
  title={X-vectors: Robust dnn embeddings for speaker recognition},
  author={Snyder, David and Garcia-Romero, Daniel and Sell, Gregory and Povey, Daniel and Khudanpur, Sanjeev},
  booktitle={2018 IEEE international conference on acoustics, speech and signal processing (ICASSP)},
  pages={5329--5333},
  year={2018},
  organization={IEEE}
}

@article{desplanques2020ecapa,
  title={Ecapa-tdnn: Emphasized channel attention, propagation and aggregation in tdnn based speaker verification},
  author={Desplanques, Brecht and Thienpondt, Jenthe and Demuynck, Kris},
  journal={arXiv preprint arXiv:2005.07143},
  year={2020}
}

@inproceedings{zhou2021resnext,
  title={Resnext and res2net structures for speaker verification},
  author={Zhou, Tianyan and Zhao, Yong and Wu, Jian},
  booktitle={2021 IEEE spoken language technology workshop (SLT)},
  pages={301--307},
  year={2021},
  organization={IEEE}
}

@article{li2017deep,
  title={Deep speaker: an end-to-end neural speaker embedding system},
  author={Li, Chao and Ma, Xiaokong and Jiang, Bing and Li, Xiangang and Zhang, Xuewei and Liu, Xiao and Cao, Ying and Kannan, Ajay and Zhu, Zhenyao},
  journal={arXiv preprint arXiv:1705.02304},
  year={2017}
}

@article{yakovlev2024reshape,
  title={Reshape dimensions network for speaker recognition},
  author={Yakovlev, Ivan and Makarov, Rostislav and Balykin, Andrei and Malov, Pavel and Okhotnikov, Anton and Torgashov, Nikita},
  journal={arXiv preprint arXiv:2407.18223},
  year={2024}
}

@article{yakovlev2026redimnet2,
  title={ReDimNet2: Scaling Speaker Verification via Time-Pooled Dimension Reshaping},
  author={Yakovlev, Ivan and Okhotnikov, Anton},
  journal={arXiv preprint arXiv:2603.11841},
  year={2026}
}

@article{chen2024eres2netv2,
  title={ERes2NetV2: Boosting short-duration speaker verification performance with computational efficiency},
  author={Chen, Yafeng and Zheng, Siqi and Wang, Hui and Cheng, Luyao and Chen, Qian and Zhang, Shiliang and Li, Junjie},
  journal={arXiv preprint arXiv:2406.02167},
  year={2024}
}

@inproceedings{kim2022rawnext,
  title={RawNeXt: Speaker verification system for variable-duration utterances with deep layer aggregation and extended dynamic scaling policies},
  author={Kim, Ju-ho and Shim, Hye-jin and Heo, Jungwoo and Yu, Ha-Jin},
  booktitle={ICASSP 2022-2022 IEEE International Conference on Acoustics, Speech and Signal Processing (ICASSP)},
  pages={7647--7651},
  year={2022},
  organization={IEEE}
}

@article{deng2026src,
  title={SRC-TDNN: Split Residual Collaborative Time Delay Neural Network for Short-Speech Speaker Verification},
  author={Deng, Fei and Deng, Lihong and Jiang, Peifan},
  journal={IEEE Signal Processing Letters},
  year={2026},
  publisher={IEEE}
}

@article{jung2020improving,
  title={Improving multi-scale aggregation using feature pyramid module for robust speaker verification of variable-duration utterances},
  author={Jung, Youngmoon and Kye, Seong Min and Choi, Yeunju and Jung, Myunghun and Kim, Hoirin},
  journal={arXiv preprint arXiv:2004.03194},
  year={2020}
}

@inproceedings{mun2023frequency,
  title={Frequency and multi-scale selective kernel attention for speaker verification},
  author={Mun, Sung Hwan and Jung, Jee-weon and Han, Min Hyun and Kim, Nam Soo},
  booktitle={2022 IEEE spoken language technology workshop (SLT)},
  pages={548--554},
  year={2023},
  organization={IEEE}
}

@inproceedings{myoung2026short,
  title={Short-Segment Speaker Verification with Pre-Trained Models and Multi-Resolution Encoder},
  author={Myoung, Jisoo and Han, Sangwook and Kim, Kihyuk and Shin, Jong Won},
  booktitle={ICASSP 2026-2026 IEEE International Conference on Acoustics, Speech and Signal Processing (ICASSP)},
  pages={18872--18876},
  year={2026},
  organization={IEEE}
}

@article{wang2018additive,
  title={Additive margin softmax for face verification},
  author={Wang, Feng and Cheng, Jian and Liu, Weiyang and Liu, Haijun},
  journal={IEEE Signal Processing Letters},
  volume={25},
  number={7},
  pages={926--930},
  year={2018},
  publisher={IEEE}
}

@inproceedings{deng2019arcface,
  title={Arcface: Additive angular margin loss for deep face recognition},
  author={Deng, Jiankang and Guo, Jia and Xue, Niannan and Zafeiriou, Stefanos},
  booktitle={Proceedings of the IEEE/CVF conference on computer vision and pattern recognition},
  pages={4690--4699},
  year={2019}
}

@inproceedings{han2023exploring,
  title={Exploring binary classification loss for speaker verification},
  author={Han, Bing and Chen, Zhengyang and Qian, Yanmin},
  booktitle={ICASSP 2023-2023 IEEE International Conference on Acoustics, Speech and Signal Processing (ICASSP)},
  pages={1--5},
  year={2023},
  organization={IEEE}
}

@inproceedings{fang2026hyperbolic,
  title={Hyperbolic Additive Margin Softmax with Hierarchical Information for Speaker Verification},
  author={Fang, Zhihua and He, Liang},
  booktitle={ICASSP 2026-2026 IEEE International Conference on Acoustics, Speech and Signal Processing (ICASSP)},
  pages={19017--19021},
  year={2026},
  organization={IEEE}
}

@inproceedings{li2026enhancing,
  title={Enhancing speaker verification with w2v-bert 2.0 and knowledge distillation guided structured pruning},
  author={Li, Ze and Cheng, Ming and Li, Ming},
  booktitle={ICASSP 2026-2026 IEEE International Conference on Acoustics, Speech and Signal Processing (ICASSP)},
  pages={16462--16466},
  year={2026},
  organization={IEEE}
}

@article{zhao2024whisper,
  title={Whisper-pmfa: Partial multi-scale feature aggregation for speaker verification using whisper models},
  author={Zhao, Yiyang and Wang, Shuai and Sun, Guangzhi and Chen, Zehua and Zhang, Chao and Xu, Mingxing and Zheng, Thomas Fang},
  journal={arXiv preprint arXiv:2408.15585},
  year={2024}
}

@inproceedings{zhang2021contrastive,
  title={Contrastive self-supervised learning for text-independent speaker verification},
  author={Zhang, Haoran and Zou, Yuexian and Wang, Helin},
  booktitle={ICASSP 2021-2021 IEEE International Conference on Acoustics, Speech and Signal Processing (ICASSP)},
  pages={6713--6717},
  year={2021},
  organization={IEEE}
}

@inproceedings{xia2021self,
  title={Self-supervised text-independent speaker verification using prototypical momentum contrastive learning},
  author={Xia, Wei and Zhang, Chunlei and Weng, Chao and Yu, Meng and Yu, Dong},
  booktitle={ICASSP 2021-2021 IEEE international conference on acoustics, speech and signal processing (ICASSP)},
  pages={6723--6727},
  year={2021},
  organization={IEEE}
}

@article{cho2022non,
  title={Non-contrastive self-supervised learning for utterance-level information extraction from speech},
  author={Cho, Jaejin and Villalba, Jes{\'u}s and Moro-Velazquez, Laureano and Dehak, Najim},
  journal={IEEE Journal of Selected Topics in Signal Processing},
  volume={16},
  number={6},
  pages={1284--1295},
  year={2022},
  publisher={IEEE}
}

@article{kingma2013auto,
  title={Auto-encoding variational bayes},
  author={Kingma, Diederik P and Welling, Max},
  journal={arXiv preprint arXiv:1312.6114},
  year={2013}
}

@article{hsu2017unsupervised,
  title={Unsupervised learning of disentangled and interpretable representations from sequential data},
  author={Hsu, Wei-Ning and Zhang, Yu and Glass, James},
  journal={Advances in neural information processing systems},
  volume={30},
  year={2017}
}

@article{kameoka2019acvae,
  title={ACVAE-VC: Non-parallel voice conversion with auxiliary classifier variational autoencoder},
  author={Kameoka, Hirokazu and Kaneko, Takuhiro and Tanaka, Kou and Hojo, Nobukatsu},
  journal={IEEE/ACM Transactions on Audio, Speech, and Language Processing},
  volume={27},
  number={9},
  pages={1432--1443},
  year={2019},
  publisher={IEEE}
}

@article{van2017neural,
  title={Neural discrete representation learning},
  author={Van Den Oord, Aaron and Vinyals, Oriol and others},
  journal={Advances in neural information processing systems},
  volume={30},
  year={2017}
}

@article{lipman2022flow,
  title={Flow matching for generative modeling},
  author={Lipman, Yaron and Chen, Ricky TQ and Ben-Hamu, Heli and Nickel, Maximilian and Le, Matt},
  journal={arXiv preprint arXiv:2210.02747},
  year={2022}
}

@article{kong2020diffwave,
  title={Diffwave: A versatile diffusion model for audio synthesis},
  author={Kong, Zhifeng and Ping, Wei and Huang, Jiaji and Zhao, Kexin and Catanzaro, Bryan},
  journal={arXiv preprint arXiv:2009.09761},
  year={2020}
}

@article{liu2023audioldm,
  title={Audioldm: Text-to-audio generation with latent diffusion models},
  author={Liu, Haohe and Chen, Zehua and Yuan, Yi and Mei, Xinhao and Liu, Xubo and Mandic, Danilo and Wang, Wenwu and Plumbley, Mark D},
  journal={arXiv preprint arXiv:2301.12503},
  year={2023}
}

@article{wang2023neural,
  title={Neural codec language models are zero-shot text to speech synthesizers},
  author={Wang, Chengyi and Chen, Sanyuan and Wu, Yu and Zhang, Ziqiang and Zhou, Long and Liu, Shujie and Chen, Zhuo and Liu, Yanqing and Wang, Huaming and Li, Jinyu and others},
  journal={arXiv preprint arXiv:2301.02111},
  year={2023}
}

@inproceedings{chen2025f5,
  title={F5-tts: A fairytaler that fakes fluent and faithful speech with flow matching},
  author={Chen, Yushen and Niu, Zhikang and Ma, Ziyang and Deng, Keqi and Wang, Chunhui and JianZhao, JianZhao and Yu, Kai and Chen, Xie},
  booktitle={Proceedings of the 63rd Annual Meeting of the Association for Computational Linguistics (Volume 1: Long Papers)},
  pages={6255--6271},
  year={2025}
}

@article{yadav2023dsvae,
  title={Dsvae: Interpretable disentangled representation for synthetic speech detection},
  author={Yadav, Amit Kumar Singh and Bhagtani, Kratika and Xiang, Ziyue and Bestagini, Paolo and Tubaro, Stefano and Delp, Edward J},
  journal={arXiv preprint arXiv:2304.03323},
  year={2023}
}

@article{bai2018empirical,
  title={An empirical evaluation of generic convolutional and recurrent networks for sequence modeling},
  author={Bai, Shaojie and Kolter, J Zico and Koltun, Vladlen},
  journal={arXiv preprint arXiv:1803.01271},
  year={2018}
}

@inproceedings{yu2024inceptionnext,
  title={Inceptionnext: When inception meets convnext},
  author={Yu, Weihao and Zhou, Pan and Yan, Shuicheng and Wang, Xinchao},
  booktitle={2024 IEEE/CVF Conference on Computer Vision and Pattern Recognition (CVPR)},
  pages={5672--5683},
  year={2024},
  organization={IEEE}
}

@article{chen2023enhanced,
  title={An enhanced res2net with local and global feature fusion for speaker verification},
  author={Chen, Yafeng and Zheng, Siqi and Wang, Hui and Cheng, Luyao and Chen, Qian and Qi, Jiajun},
  journal={arXiv preprint arXiv:2305.12838},
  year={2023}
}

@article{wang2023cam++,
  title={Cam++: A fast and efficient network for speaker verification using context-aware masking},
  author={Wang, Hui and Zheng, Siqi and Chen, Yafeng and Cheng, Luyao and Chen, Qian},
  journal={arXiv preprint arXiv:2303.00332},
  year={2023}
}

@inproceedings{snyder2017deep,
  title={Deep neural network embeddings for text-independent speaker verification},
  author={Snyder, David and Garcia-Romero, Daniel and Povey, Daniel and Khudanpur, Sanjeev},
  booktitle={Proc. Interspeech 2017},
  pages={999--1003},
  year={2017}
}

@article{chung2018voxceleb2,
  title={Voxceleb2: Deep speaker recognition},
  author={Chung, Joon Son and Nagrani, Arsha and Zisserman, Andrew},
  journal={arXiv preprint arXiv:1806.05622},
  year={2018}
}

@article{nagrani2020voxceleb,
  title={Voxceleb: Large-scale speaker verification in the wild},
  author={Nagrani, Arsha and Chung, Joon Son and Xie, Weidi and Zisserman, Andrew},
  journal={Computer Speech \& Language},
  volume={60},
  pages={101027},
  year={2020},
  publisher={Elsevier}
}

@inproceedings{fan2020cn,
  title={Cn-celeb: a challenging chinese speaker recognition dataset},
  author={Fan, Yue and Kang, JW and Li, LT and Li, KC and Chen, HL and Cheng, ST and Zhang, PY and Zhou, ZY and Cai, YQ and Wang, Dong},
  booktitle={ICASSP 2020-2020 IEEE International Conference on Acoustics, Speech and Signal Processing (ICASSP)},
  pages={7604--7608},
  year={2020},
  organization={IEEE}
}

@article{chung2020spot,
  title={Spot the conversation: speaker diarisation in the wild},
  author={Chung, Joon Son and Huh, Jaesung and Nagrani, Arsha and Afouras, Triantafyllos and Zisserman, Andrew},
  journal={arXiv preprint arXiv:2007.01216},
  year={2020}
}

@inproceedings{yu2022m2met,
  title={M2MeT: The ICASSP 2022 multi-channel multi-party meeting transcription challenge},
  author={Yu, Fan and Zhang, Shiliang and Fu, Yihui and Xie, Lei and Zheng, Siqi and Du, Zhihao and Huang, Weilong and Guo, Pengcheng and Yan, Zhijie and Ma, Bin and others},
  booktitle={ICASSP 2022-2022 IEEE International Conference on Acoustics, Speech and Signal Processing (ICASSP)},
  pages={6167--6171},
  year={2022},
  organization={IEEE}
}

@article{fu2021aishell,
  title={Aishell-4: An open source dataset for speech enhancement, separation, recognition and speaker diarization in conference scenario},
  author={Fu, Yihui and Cheng, Luyao and Lv, Shubo and Jv, Yukai and Kong, Yuxiang and Chen, Zhuo and Hu, Yanxin and Xie, Lei and Wu, Jian and Bu, Hui and others},
  journal={arXiv preprint arXiv:2104.03603},
  year={2021}
}

@inproceedings{carletta2005ami,
  title={The AMI meeting corpus: A pre-announcemFent},
  author={Carletta, Jean and Ashby, Simone and Bourban, Sebastien and Flynn, Mike and Guillemot, Mael and Hain, Thomas and Kadlec, Jaroslav and Karaiskos, Vasilis and Kraaij, Wessel and Kronenthal, Melissa and others},
  booktitle={International workshop on machine learning for multimodal interaction},
  pages={28--39},
  year={2005},
  organization={Springer}
}

@inproceedings{chen20253d,
  title={3d-speaker-toolkit: An open-source toolkit for multimodal speaker verification and diarization},
  author={Chen, Yafeng and Zheng, Siqi and Wang, Hui and Cheng, Luyao and Zhu, Tinglong and Huang, Rongjie and Deng, Chong and Chen, Qian and Zhang, Shiliang and Wang, Wen and others},
  booktitle={ICASSP 2025-2025 IEEE International Conference on Acoustics, Speech and Signal Processing (ICASSP)},
  pages={1--5},
  year={2025},
  organization={IEEE}
}

@article{wang2024advancing,
  title={Advancing speaker embedding learning: Wespeaker toolkit for research and production},
  author={Wang, Shuai and Chen, Zhengyang and Han, Bing and Wang, Hongji and Liang, Chengdong and Zhang, Binbin and Xiang, Xu and Ding, Wen and Rohdin, Johan and Silnova, Anna and others},
  journal={Speech Communication},
  volume={162},
  pages={103104},
  year={2024},
  publisher={Elsevier}
}

@article{mcinnes2018umap,
  title={Umap: Uniform manifold approximation and projection for dimension reduction},
  author={McInnes, Leland and Healy, John and Melville, James},
  journal={arXiv preprint arXiv:1802.03426},
  year={2018}
}

@inproceedings{peebles2023scalable,
  title={Scalable diffusion models with transformers},
  author={Peebles, William and Xie, Saining},
  booktitle={2023 IEEE/CVF International Conference on Computer Vision (ICCV)},
  pages={4172--4182},
  year={2023},
  organization={IEEE}
}

@article{beltagy2020longformer,
  title={Longformer: The long-document transformer},
  author={Beltagy, Iz and Peters, Matthew E and Cohan, Arman},
  journal={arXiv preprint arXiv:2004.05150},
  year={2020}
}

@inproceedings{bredin2023pyannote,
  title={pyannote. audio 2.1 speaker diarization pipeline: principle, benchmark, and recipe},
  author={Bredin, Herv{\'e}},
  booktitle={24th Interspeech Conference (INTERSPEECH 2023)},
  pages={1983--1987},
  year={2023},
  organization={ISCA}
}

@inproceedings{yu2021cam,
  title={Cam: Context-aware masking for robust speaker verification},
  author={Yu, Ya-Qi and Zheng, Siqi and Suo, Hongbin and Lei, Yun and Li, Wu-Jun},
  booktitle={ICASSP 2021-2021 IEEE International Conference on Acoustics, Speech and Signal Processing (ICASSP)},
  pages={6703--6707},
  year={2021},
  organization={IEEE}
}

@inproceedings{radford2023robust,
  title={Robust speech recognition via large-scale weak supervision},
  author={Radford, Alec and Kim, Jong Wook and Xu, Tao and Brockman, Greg and McLeavey, Christine and Sutskever, Ilya},
  booktitle={International conference on machine learning},
  pages={28492--28518},
  year={2023},
  organization={PMLR}
}

@inproceedings{han2025leveraging,
  title={Leveraging self-supervised learning for speaker diarization},
  author={Han, Jiangyu and Landini, Federico and Rohdin, Johan and Silnova, Anna and Diez, Mireia and Burget, Luk{\'a}{\v{s}}},
  booktitle={ICASSP 2025-2025 IEEE International Conference on Acoustics, Speech and Signal Processing (ICASSP)},
  pages={1--5},
  year={2025},
  organization={IEEE}
}

@inproceedings{yin2026whisperdiari,
  title={WhisperDiari: A Whisper-Based Speaker Diarization Framework in Token Space Leveraging Semantic and Speaker Information for Better Text Adaptability},
  author={Yin, Yongkang and Zou, Yuexian},
  booktitle={Proceedings of the AAAI Conference on Artificial Intelligence},
  volume={40},
  number={41},
  pages={34477--34485},
  year={2026}
}

@inproceedings{vaessen2022fine,
  title={Fine-tuning wav2vec2 for speaker recognition},
  author={Vaessen, Nik and Van Leeuwen, David A},
  booktitle={ICASSP 2022-2022 IEEE International Conference on Acoustics, Speech and Signal Processing (ICASSP)},
  pages={7967--7971},
  year={2022},
  organization={IEEE}
}

@article{ho2020denoising,
  title={Denoising diffusion probabilistic models},
  author={Ho, Jonathan and Jain, Ajay and Abbeel, Pieter},
  journal={Advances in neural information processing systems},
  volume={33},
  pages={6840--6851},
  year={2020}
}

@inproceedings{zhu2026omnihuman,
  title={OmniHuman: a large-scale dataset and benchmark for human-centric video generation},
  author={Zhu, Lei and Cai, Xing and Chen, Yingjie and Li, Yiheng and Yang, Binxin and Liu, Hao and Chen, Jie and Li, Chen and LYu, Jing},
  booktitle={European Conference on Computer Vision},
  pages={92--110},
  year={2026},
  organization={Springer}
}

\vfill

\end{document}